\documentclass[11pt]{amsart}
\usepackage{amssymb,amsmath,amsfonts,amsaddr}
\usepackage{graphicx,psfrag,epsfig,color}
\usepackage{setspace}
\usepackage{subfigure}
\usepackage{bm}
\usepackage[normalem]{ulem}
\usepackage[left=1.0in, right=1.0in, top=1.0in, bottom=1.0in,
  includehead,includefoot]{geometry}

\usepackage{hyperref}

\newcommand{\bit}{\begin{itemize}}
\newcommand{\eit}{\end{itemize}}
\newcommand{\beq}{\begin{equation}}
\newcommand{\eeq}{\end{equation}}
\newcommand{\ben}{\begin{enumerate}}
\newcommand{\een}{\end{enumerate}}
\newcommand{\bal}{\begin{align}}
\newcommand{\eal}{\end{align}}

\numberwithin{equation}{section}
\begin{document}

\title[Rac-Rho-ECM dynamics and cell shape I]{Cellular tango: How extracellular matrix adhesion choreographs Rac-Rho signaling and cell movement}

\author{Elisabeth G. Rens}
\address{Current: Delft Institute of Applied Mathematics, Delft University of Technology, Delft, the Netherlands, Previous: Department of Mathematics, University of British Columbia, Vancouver, Canada}

\author{Leah Edelstein-Keshet}
\address{Department of Mathematics, University of British Columbia, Vancouver, Canada}
\date{\today}

\maketitle

\begin{abstract}
The small GTPases Rac and Rho are known to regulate eukaryotic cell shape, promoting front protrusion (Rac) or rear retraction (Rho) of the cell edge. Such cell deformation changes the contact and adhesion of cell to the extracellular matrix (ECM), while ECM signaling through integrin receptors also affects GTPase activity. We develop and investigate a model for this three-way feedback loop in 1D and 2D spatial domains, as well as in a fully deforming 2D cell shape. The model consists of reaction-diffusion equations solved numerically with open-source software, Morpheus, and with custom-built cellular Potts model simulations. We find a variety of patterns and cell behaviors, including persistent polarity,  flipped front-back cell polarity oscillations, and random protrusion-retraction. 
We show that the observed spatial patterns depend on the cell shape, and vice versa. The cell stiffness and biophysical properties also affect patterning, overall cell migration phenotypes.
    
\end{abstract}

%\tableofcontents

\section{Introduction}

 Cell migration is a vital phenomenon that occurs in health and disease, including
 wound healing and cancer metastasis. Signals that  promote cell migration include chemical gradients, topographic or mechanical cues or adhesion gradients \cite{Houk2012,SahaiECM2014,park2016}. Properties of a cell's internal regulatory system, and the various aspects of its environment, together determine the spatiotemporal distribution of intracellular signaling, and the resultant cell response \cite{ridley2003}. The onset, direction, persistence and type of cell migration has important biological significance. During wound healing, cells need to start moving in the right direction to properly close a wound. In cancer, spontaneous persistent migration is a major determinant of metastasis. Cell mutations and cell response to changes in the environmental cues affect whether cell migration is normal, as in wound healing and development, or defective, as in cancer metastasis.
 
 Cell motility is regulated by signaling networks in the cell, where proteins of the Rho-family GTPases, called Rac and Rho are central hubs \cite{Ridley-01,etienne2002, burridge2004,Etienne-04}. These proteins can exist in an activated, membrane bound form, or inactive form that freely moves in the cytosol \cite{Bourne1990}. Their (in)activation is regulated by GEF/GAP proteins. Rac is generally localized in protrusions, where it promotes actin assembly to further push the cell membrane out \cite{burridge2006}. Rho is generally activated at the trailing ends of the cell, where it promotes cell contraction via ROCK, a kinase that recruits and activates  myosin on the actin cytoskeleton \cite{maekawa1999,Machacek-09}. Together, these two opposing actions of Rac and Rho on the cytoskeleton regulates the location of protrusive or %\sout{retractive}
contractile activities \cite{guilluy2011,Parri-10,Sailem-14}. 
 
 One typical, and relatively well-studied, example of a Rac/Rho pattern is two opposing gradients, consistent with persistently polarized cell motion \cite{ridley2003}. However, the spatial distribution of Rac/Rho is not always this simple. Complex spatial and temporal patterning have been observed. For instance, Rho exhibits transient bursts of activity at the trailing edge \cite{pertz2006}. Also, waves of Rho from the front to the back of the cell have been observed \cite{Bolado-C-2020}. The activity at multiple sites of protrusion is often associated with high Rac, but bands of intense Rho activity have been also been observed in active protrusions \cite{pertz2006}. It is suggested that Rac and Rho actually cyclically interchange their activity in protrusions \cite{tomar2009}. All this shows that the dynamics and distribution of Rac-Rho are non-trivial and depend on the exact internal signaling network, cell geometry and environmental inputs.

Mutations can affect protein conformations, and their binding and activation rates, and this could bias the competition between Rac and Rho and thus the expected cell behavior. Signals from the environment also play a significant role. Cells are surrounded by the extracellular matrix (ECM), a network of proteins and fibers \cite{Sheetz1998}. The mechanical architecture of the ECM, a topic of increasing interest in cancer and wound healing, is an important determining factor for cell
phenotype \cite{SahaiECM2014}. Signals from outside of the cell, such as external forces \cite{lessey2012,poh2009}, cell-ECM bonds \cite{Lee2020,Plotnikov2012} or growth factors \cite{Aoki-05}, are channeled to GTPases like Rac, Rho and Cdc42, and promote or inhibit their activation or inactivation. Cell protrusions interact with the ECM, causing external cues to change local GTPase activity, which by diffusion also affects the signaling profile of the whole cell. 

The interplay between ECM and cell signaling has received much attention in recent years.
For instance, one study \cite{Lee2020} looked at cell migration on patterned adhesive islands. Their experiments show that the precise local pattern determines where Cdc42 was activated. The authors went on to show that with the right adhesive pattern, cell migration could be reversed. Another study used different sizes of and spacing between ECM islands to show that the onset of Rac waves correlated with local adhesion, affecting the cells' orientation of motion \cite{xia2008}.

Various mathematical models have considered GTPase dynamics within a cell but most of these studies only focused on cell polarization, such as \cite{Maree-06,Jilkine-11,Holmes_PLoS_12,kopfer2020,wang2017,tao2020,cao2019,camley2017,Zmurchok2020b,copos2020} and others. The role of cell-ECM adhesion has been mathematically modelled and studied in detail in the context of cell spreading and directional migration (see e.g. \cite{Rens2020,Deshpande2008,Ronan2014,Vernerey2014,bangasser2017} and many others), but not so much towards GTPase patterning \cite{hastings2019}. So, what still remains elusive is how various, more diverse phenotypes in Rac-Rho patterning can emerge. Here we focus specifically on the patterning in the context of ECM signaling.

Even though there are many complex interactions between a cell and the ECM, it was proposed that experimental observations of melanoma cells  \cite{JSpark2017} can be explained by a minimal model of Rac-Rho-ECM  \cite{JSpark2017, Holmes2017} where adhesion promotes Rho activation, Rac and Rho are mutually antagonistic,  Rac upregulates adhesion and Rho down-regulates adhesion. Both these papers motivate our own work. There, the  dynamics were studied in a 2-compartment model, representing the front and the back of a cell.

In this paper, we will explore the effect of this ECM feedback mechanism on the full spatiotemporal dynamics of Rac and Rho. We focus our plan on addressing the following questions (1)  Can we find the same dynamic cell polarity patterns as described in \cite{JSpark2017} in the full spatial models? What are minimal requirements to get the oscillatory patterns of behaviour observed in melanoma cells in \cite{JSpark2017}?  (2) 
Do realistic force-dependent adhesion dynamics give rise to the same patterning as in \cite{JSpark2017}? (3) What are the possible GTPase patterning dynamics in the spatial model? How does the geometry affect those dynamics? 
(4) How does the feedback between GTPase activity and cell protrusion/retraction affect the resulting pattern dynamics and the cell behaviour? 
(5) How do biophysical properties such as cell stiffness affect the patterns and resulting cell phenotype?

 \begin{figure}
     \centering
       \includegraphics[scale=0.7]{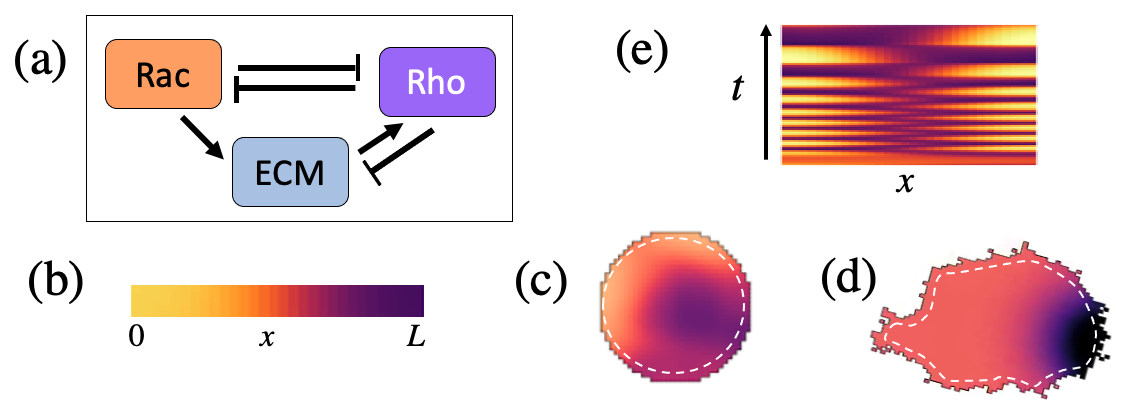}
     \caption{A schematic diagram of the models considered in this paper. (a) The well-mixed mutual inhibition model, %Eqs.~\eqref{eq:WellMixedRacRho}, 
     for Rac -Rho without and later with feedback from the extracellular matrix (ECM). (b)-(d) spatially distributed model variants where PDEs describe the basic interactions distributed across the cell. Rac and Rho then have both active and inactive forms that diffuse at distinct rates. ECM signaling is distributed spatially as well. 
     The domain geometry considered is (b) a 1D strip, (c) a static circular domain and 
     (d) a full 2D version of the model with cell shape captured by a cellular Potts computation. In all three cases, we represent the results in a kymograph, with the spatial variable horizontal. For 2D geometries, the spatial variable is the perimeter of the full 2D domain and $-\pi \le x \le \pi$.}
     \label{fig:SchematicOfModels}
 \end{figure}

%%%%%%%%%%%%%%%%%%%

\subsection{The extracellular matrix (ECM)}

The ECM is a  meshwork of fibrous proteins such as collagen surrounding cells. It presents a complex topography and adhesive environment on which cells crawl, pull, deform, and remodel \cite{Sheetz1998,Larsen2006}. As cells pull and exert forces on the ECM, the resulting mechanical tension creates stimuli that affect the cells.

\begin{figure}
    \centering
    \includegraphics[scale=1]{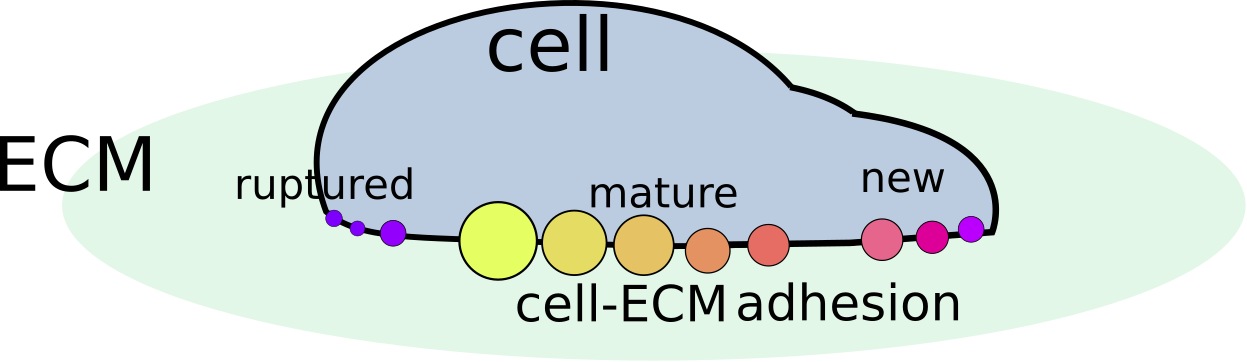}
    \caption{Side view of cell with new cell-ECM adhesion bonds forming in front and old ones rupturing in back}
    \label{fig:cellschematic}
\end{figure}

 Cells attach to ECM via integrin bonds, that are distributed over the cell surface. A migrating cell creates new integrin attachments to the ECM as the cell front expands to new regions. The rear of the cell detaches, breaking some of the existing bonds. See Figure \ref{fig:cellschematic} for a cartoon that illustrates this concept. Engagement of integrin receptors then signals to GTPases \cite{costa2013,huveneers2009,Lawson2014}. Experimental evidence suggests that ECM signaling strongly enhances the activation of RhoA \cite{Danen2002,Park2011}, although it likely also have positive effects on Rac activity \cite{arthur2000,costa2013}.
 
\subsection{Integrin bonds}

Integrins cluster into focal adhesions (FA's), creating local hot-spots of adhesion. Structural proteins within FAs enable binding of the integrins to the cytoskeleton, which stabilize the adhesion and allows cells to transmit cytoskeletal forces to the integrins. The biophysical properties of integrins have been elucidated in recent years \cite{Kong2009}. The  binding-unbinding kinetics of integrin bonds are influenced by mechanical force. For a long time, it was thought that the detachment rate of integrins increases with force (slip-bond), but recent experiments revealed that certain integrins \cite{Kong2009} actually behave like catch-bonds. Applying force to a catch-bond causes tightening and reduces the rate of unbinding up to some force threshold, beyond which the bonds start to break. Hence, the lifetime of a catch-bond is non-monotonic, and maximal for some intermediate force magnitude. 

Experimental \cite{mackay2020} and modeling  (\cite{Nicolas2005,Besser2006,mackay2019} and many others) papers have investigated the relationship between applied force and adhesion bond growth and breakage. In this work, we followed the ideas of Novikova \& Storm \cite{Novikova2013}, who approximated a focal adhesion as a cluster of catch-slip bonds of which the cluster size is optimal under finite force (catch) and ruptures above a certain force threshold (slip). In this paper, we compare this model with a pure slip-bond model. 
 
 In our ultimate deforming cell model, we keep track of parts of the cell that advances, forming fresh integrin bonds and establishing ECM signaling.
 Retraction of parts of the cell result in forces pulling on these adhesion bonds, and tearing them off, which is also included in our ultimate cell deformation simulations.

  %%%%%%%%%%%%%%%%%%%%%%%%%%%%%%
 \subsection{Experimentally observed phenotypes} 
Much of the work reported here was motivated by the experiments of \cite{JSpark2017} on aggressive melanoma cells (1205Lu) adhering to topographical surfaces mimicking ECM. These substrates were arrays of nano-posts of varying densities and anisotropy, coated with the adhesion protein fibronectin (FN). Our work here follows on the foundations built by the theoretical work on \cite{Holmes2017}. We briefly summarize the key experimental observations.

The cells in \cite{JSpark2017} expressed fluorescent tagged pleckstrin homology (PH) domain of Akt, an indicator of the activity of PI3K, a kinase associated with the activity of Rac at the ``cell front''. This region corresponded with protruding lamellipodia. The location of the cell front was tracked over time under various conditions. The dynamics were classified into three main types, persistent, random, or oscillatory. 
Persistent cells had a stable ``front'', and tended to be polarized. In oscillatory cells, the PI3K ``hot spot'' switched back and forth across the cell, so that front and back polarity exchanged in some rhythmic process. In the ``random cells'', the region of ``frontness'' jumped from one site to another along the cell edge, with no clear periodic pattern. 
 
 In \cite{JSpark2017}, every experiment produced some fraction of each of the three types, with relative proportions dependent on the type of manipulation. 
 The adhesion of the cell and its access to the ECM could be changed by modifying post density.
 On sparser posts arrays, cells penetrate between the posts and attach to the underlying ECM. Higher post arrays (hence lower cell-ECM contact) led to a greater proportion of random cells and less persistently polarized cells \cite{JSpark2017}.

The model in \cite{Holmes2017} could account for major experimental observations in \cite{JSpark2017} with a minimal model based on Rac-Rho mutual inhibition, opposed effects of Rac and Rho on cell expansion (Rac) and contraction (Rho), together with feedback from the ECM to the GTPases. Our paper proceeds to explore this idea with greater spatial resolution and with enhanced computational tools.

%%%%%%%%%%%%%%%
\section{Modeling background}
We briefly summarize background information for the GTPase models, assumptions, and common formulation on which this paper rests. Details for these models can be found in \cite{Holmes-15,Holmes2016}. 

Each GTPase acts like a ``molecular switch'', where only the active ``ON'' state, bound to the plasma membrane, has downstream influence. However, limited availability of inactive GTPase can affect the rate of activation. On the timescale of interest (seconds, minutes), the total amount of a given GTPase is roughly constant in the cell. Thus, for GTPase $j$, a pair of partial differential equations (PDEs) is used to keep track of the distribution of active ($G^j$) and inactive ($G_I^j$) forms:
\begin{subequations}\label{eq:genericG,G_i,E}
\begin{align}
\frac{dG^j}{dt}&= A_j G_I^j - I_j G^j + D^j \triangle G^j,\\   
\frac{d{G_I}^j}{dt}&= -A_j G_I^j + I_j G^j + D_I^j \triangle G_I^j.
\end{align}
Here $A_j, I_j$ are rates of GTPase activation and inactivation, that  depend on interactions with other GTPases and the ECM. Inactive GTPase resides in the cytosol, and diffuses faster than active GTPase, so that $D^j \ll D_I^j$.

We also track the adhesion of the cell to the ECM. A typical equation for this ECM adhesion variable is,
\begin{equation}\label{eq:generic_E}
\frac{dE}{dt}= \epsilon(a-d E) .  
\end{equation}
\end{subequations}
 The ECM rate of increase, $a$, and decay $d$ will be important GTPase-dependent terms describe below, and $\epsilon$ is a  small parameter signifying slower dynamics. The ECM does not diffuse.  The set of equations \eqref{eq:genericG,G_i,E} for the case of Rac ($R$) and Rho ($\rho$) lead to a total of five PDES. The specific assumptions are described next.

\subsection{Rac-Rho mutual antagonism}
For the mutually antagonistic Rac and Rho, we follow the
basic assumptions in \cite{Holmes2016,JSpark2017,Holmes2017} to arrive at the GTPase equations
\begin{subequations}\label{eq:BasicRac-RhoSys}
\begin{align}
\frac{dR}{dt}&=A_R(\rho) R_I-I_R R + D \triangle R, \qquad \frac{dR_I}{dt}=-A_R(\rho) R_I+I_R R + D_I \triangle R_I,\\
\frac{d\rho}{dt}&=A_\rho(R,E) \rho_I-I_\rho \rho + D \triangle \rho, 
\qquad\frac{d\rho_I}{dt}=-A_\rho(R,E) \rho_I+I_\rho \rho + D_I \triangle \rho_I,
\end{align}
where
\begin{equation}\label{eq:A_RI_R}
 A_R(\rho)= \frac{b_R}{1+\rho^m}, \quad
 A_\rho(R,E)= \frac{b_\rho(E)}{1+R^m}, \quad I_R=I_\rho = \delta.
\end{equation}
\end{subequations}
The crosstalk of Rac and Rho is represented in the activation rates, whereas inactivation rates have been taken as constant.

%%%%%%%%%%%%%%%%%%%%%
\subsection{Rac-Rho ECM feedback}

We add two-way feedback from ECM to Rho and from Rac and Rho to the ECM by assuming that
\begin{subequations}
\label{eqn:ECM_01.1}
\begin{align}
\frac{dE}{dt}&= \epsilon(a(R,\rho,E)-d(R,\rho,E) E) ,\\
\mbox{and  }\quad b_\rho(E)&= %\gamma_E f_{E}(E)=
k_E+ \gamma_E \frac{E^m}{(E_0^m+E^m)}.
\end{align}
\end{subequations}
The Rho activation rate $b_\rho(E)$ is affected by ECM signaling  \cite{Danen2002,Park2011}.
In \eqref{eqn:ECM_01.1}b, $k_E$, is a basal Rho activation rate, and $\gamma_E$ is a parameter governing the strength of feedback from 
ECM signals to Rho activation. %

The basic assumptions made in Eqs.~\eqref{eq:A_RI_R} and \eqref{eqn:ECM_01.1}b will be a common basis for all models discussed in this paper. Details of the assumptions for the ECM equation \eqref{eqn:ECM_01.1}a will be discussed in what follows.

\section{Model variants}

The three model variants we consider differ only in the way that the ECM reacts to the downstream effects of Rac and Rho. In all three cases, we will see that Rac feedback is positive, and Rho feedback is negative, but the latter two models are formulated with a greater detail about integrin bond formation and breakage.

\subsection{Model I: the original version}

In the original model \cite{JSpark2017, Holmes2017}, it was reasoned that cell-ECM contact increases when Rac causes the cell to spread, and decreases when Rho causes cell contraction. This idea was modeled with a assumptions for \eqref{eq:generic_E} that: 
\begin{equation} \label{eq:ECMstuffForModelI}
a= a(R)=K+\gamma_R \frac{R^n}{(R_0^n+R^n)},\quad d =d(\rho)= k_P+ \gamma_\rho  \frac{\rho^n}{(\rho_0^n+\rho^n)}.
\end{equation}
We refer to this version as Model I.

\subsection{Integrin bonds and ECM dynamics}

In the integrin bond model variants, we take into account a force $F(R,\rho)$ exerted on adhesion bonds with basic form
\begin{subequations}\label{eq:ECMandForces}
\begin{equation}\label{eq:forceonbond}
F(R,\rho)=\beta_\rho \frac{\rho}{1+\rho}-\beta_R \frac{R}{1+R},
\end{equation}
and a lower bound of 0. Rho-driven contraction pulls on adhesion bonds. When Rac dominates and leads to local protrusion, there is no force on the bonds, given that the cell ``rolls over'' those adherent sites.
The ECM equation is then taken to be 
\begin{equation}\label{eq:ECM_ModelII_III}
\frac{dE}{dt}=\epsilon \left(K(E_t-E) -  d(F,E) E \right).
\end{equation}
\end{subequations}
So, $a=a(E)=K(E_t-E)$ and $d(F,E)$ is the force-dependent integrin bond breakage rate, where the amount of force depends on Rac and Rho. In the absence of force, the ECM variable settles into its steady state level, $E_t$, on a timescale of $1/(\epsilon K)$. We can interpret $E_t$ as the maximal local density of bound integrin bonds. Rac and Rho activity creates force that affects the bond breakage, and impacts ECM signaling. To arrive at reasonable assumptions about $d(F,E)$ we used the integrin biophysics model of \cite{Novikova2013}.

\begin{figure}
    \centering
    \includegraphics[width=0.8\textwidth]{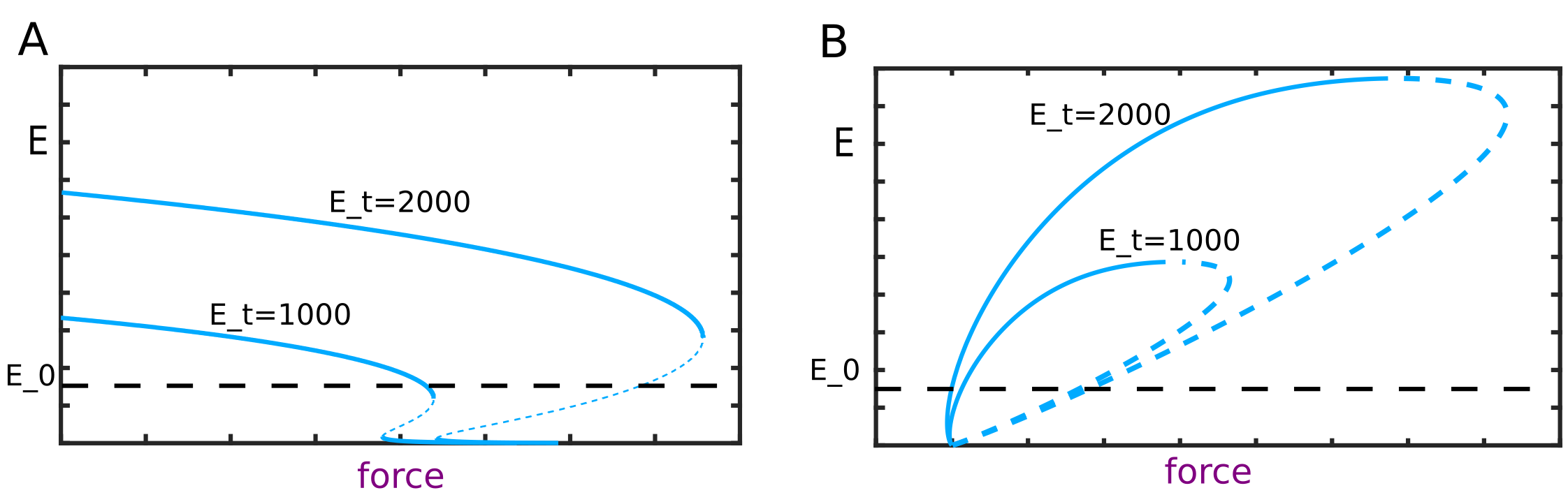}
    \caption{Bifurcation diagrams for \eqref{eq:ECM_ModelII_III} showing the steady state level of adherent integrin bonds, $E$, as a function of applied force, $F$. Solid lines indicate the stable branches and dotted lines are the unstable branches. $E_0$ is the level of adhesion at which Rho activation turns on, see \eqref{eqn:ECM_01.1}. (A) Steady-state $E$ for slip-bond case, with $d(F,E)$ as in \eqref{eq:db_slip}. Higher force leads to lower adhesion. (B) as in (A) but for the catch-slip-bond, $d(F,E)$ given by \eqref{eq:dbCatchSplip}. Higher force leads to higher adhesion, until the force reaches a maximum and adhesions break down completely. In each case, two values of the maximal adhesion $E_t$ are shown. }
    \label{fig:steadystates-slip-vs-catch}
\end{figure}

\subsubsection{Model II: Slip bond}
The rate of bond dissociation of slip-bonds is well-approximated by
\begin{equation}\label{eq:db_slip}
d(F,E)= k_0\exp \left(\frac{F}{p(E+E_s)}\right).
\end{equation}
The total force is distributed over all local integrin bonds at a given adhesion site, and $F/E$ is a force per bond that leads to rupture. The small correction $E_s$ in the denominator prevent blowup as $E \to 0$, ensuring that small bonds under low force can
grow.The parameter $p$ is a reference value of force per bond, making the term in large braces dimensionless.
When $F=0$, the bond breakage rate is $d=k_0$, whereas when $F/E \approx p$, the breakage rate is $d \approx k_0 e^1 \approx 2.7 k_0$. Hence $p$ sets the reference force per bond at which adhesions detach at 2.7 times their basal rate of detachment. Larger $p$ means that greater force per bond is needed to cause the same bond breakage rate.

\subsubsection{Model III: Catch-slip bond} 
In this case, we take
\begin{equation}
\label{eq:dbCatchSplip}
d= k_0\exp \left(\frac{F}{p(E+E_s)}\right)+k_{0c}\exp \left(-\frac{F}{p(E+E_s)}\right).
\end{equation}
Catch-bonds tend to grow stronger under the influence of  force up to some limit, so that their lifetime is maximal under some optimal force. %(see Figure \ref{fig:oscillations-explained}B).
Beyond that point, for larger applied force, the bonds break. We adopt the values $k_0 \approx 0.0004,k_{0c} \approx 55$, appropriate to $\alpha 5-\beta 1$ integrin \cite{Novikova2013}.

\subsubsection{ECM-force bifurcation plot}
To gain some intuition, we compared the force-dependent adhesion for catch-slip versus slip-bonds. To do so, we isolated Eqn.~\eqref{eq:ECM_ModelII_III}, together with either of the two force-dependent bond dissociation rates
\eqref{eq:db_slip} or 
\eqref{eq:dbCatchSplip} (leaving out the Rac-Rho dependence of force). We 
plotted the steady state adhesion $E$ as a function of force $F$ in Figure~\ref{fig:steadystates-slip-vs-catch}. We also compare the value $E_0$ (level of adhesion at which ECM-signaling to Rho turns on in Eqn.~\eqref{eqn:ECM_01.1}) on the same plots. 
Figure~\ref{fig:steadystates-slip-vs-catch} shows that in the case of a slip-bond, the adhesion cluster decreases as force increases and quickly ruptures above some force threshold, implying that (in the full model) signaling to Rho is only sustained at low force. In the case of catch-slip bonds, the adhesion clusters increase as force increase up to some threshold for rupture. In that case,  signaling to Rho is most strongest at intermediate force magnitude. When there is more ECM contact (compare $E_t=2000$ with $E_t=1000$), the clusters are larger and more long-lived, as more force is required for rupturing the bonds.

\subsection{Model summary}
Briefly, all three models use the basic Rac-Rho equations \eqref{eq:BasicRac-RhoSys} and \eqref{eqn:ECM_01.1}, with a generic ECM dynamics. 
Model II assumes slip bond dynamics, and Model III takes slip-catch bond dynamics for the ECM equation. The three variants are summarized in Table~\ref{tab:ModelSummary}.

\begin{table}[htbp]
    \centering
    \begin{tabular}{|l |l| l| l|} \hline
     Model Type & Description  &Equations  & Ref\\ \hline \hline
     Original (I)  &Generic ECM eqn.&
     \eqref{eq:BasicRac-RhoSys}, \eqref{eqn:ECM_01.1},  \eqref{eq:ECMstuffForModelI} & \cite{JSpark2017,Holmes2017}\\
    Integrin bonds (II)   &Slip-bond &\eqref{eq:BasicRac-RhoSys},\eqref{eqn:ECM_01.1}b,\eqref{eq:ECMandForces}, \eqref{eq:db_slip} & \cite{Novikova2013} \\
     Integrin bonds (III)& Slip-catch bond &\eqref{eq:BasicRac-RhoSys},\eqref{eqn:ECM_01.1}b,\eqref{eq:ECMandForces}, \eqref{eq:dbCatchSplip}  & \cite{Novikova2013} \\  \hline
    \end{tabular}
    \caption{Summary of the three models considered in this paper. All models use the basic set of Rac-Rho equations, \eqref{eq:BasicRac-RhoSys}, but the specific assumptions about the ECM dynamics differ, as indicated in this table.}
    \label{tab:ModelSummary}
\end{table}

\section{Preliminary results: spatially uniform case}

There are several submodels and simplified cases that are of interest. We briefly survey some previous results and link to the fuller spatial models.

\subsection{Well-mixed Rac-Rho model}

\begin{figure}
    \centering
    \includegraphics[scale=0.65]{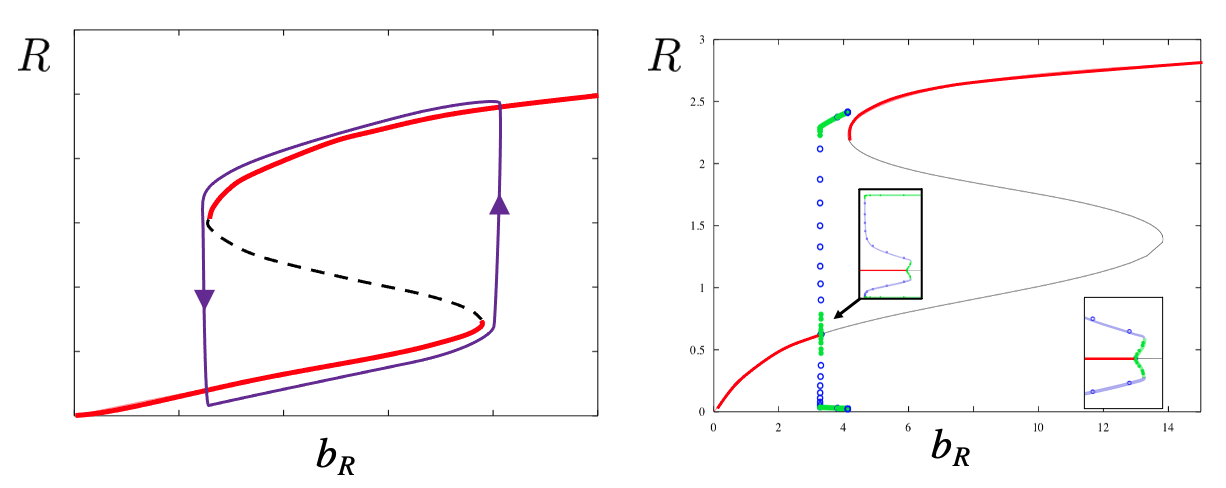}
    \caption{Left: The well-mixed Rac-Rho system \eqref{eq:WellMixedRacRho2} on its own is bistable, leading to an S-shaped bifurcation diagram with respect to a parameter such as $b_R$ or $b_\rho$ (not shown). Slow negative feedback that shifts a parameter (for example, the effect of the ECM on $b_\rho$) can, in principle, lead to cycling around the hysteresis loop, and emergence of limit cycle oscillations. Right: The well-mixed Rac-Rho-ECM (Model I) bifurcation diagram. Some of the Rac-Rho bifurcation structure is preserved, but we see a Hopf bifurcation at some intermediate value of the parameter $b_R$. There is a very narrow range of $b_R$ where two stable limit cycles coexist, as shown in the two zoom insets. 
    }
    \label{fig:RacRho2BifurcDiag}
\end{figure}

 A simple reduced variant of the model for only Rac and Rho (with no ECM,  $b_\rho$= constant, and no spatial terms) is the well-mixed version of \eqref{eq:BasicRac-RhoSys} where spatial gradients are ignored. In that case, it is common to assume that the total GTPase of each type, $G_T= R_T, \rho_T$ is constant in the cell on the timescale of interest, so that %the amount of the inactive GTPase satisfies
\[
R_T=R+R_I,\quad \rho_T=\rho+\rho_I.
\]
Hence, the inactive forms 
can be eliminated from the system, resulting in a pair of ordinary differential equations (ODEs), \eqref{eq:WellMixedRacRho2}. This mutual-antagonism system is  
is bistable for some range of parameter values \cite{Holmes2016}, as shown in the left panel of Figure~\ref{fig:RacRho2BifurcDiag}. Bistability implies hysteresis, and slow negative feedback  from some influence can lead to an excursion around the hysteresis loop that results in oscillations. The ECM in the Rac-Rho-ECM model plays the role of this additional feedback. 

%%%%%%%%%%%%%%%%%%%%%%%%%%%

\subsection{Well-mixed Rac-Rho-ECM model (I)}

Once ECM is coupled to the Rac-Rho system in for Model I, interesting dynamics emerge. For example, the  well-mixed variant of the Rac-Rho-ECM Model I,  whose equations are given by \eqref{eq:LEK_ModelI} has a regime of oscillations sandwiched between states of high and low Rac. A typical bifurcation diagram for \eqref{eq:LEK_ModelI} is shown on the right panel of Figure~\ref{fig:RacRho2BifurcDiag}.

\begin{figure}
    \centering
    \includegraphics[scale=0.5]{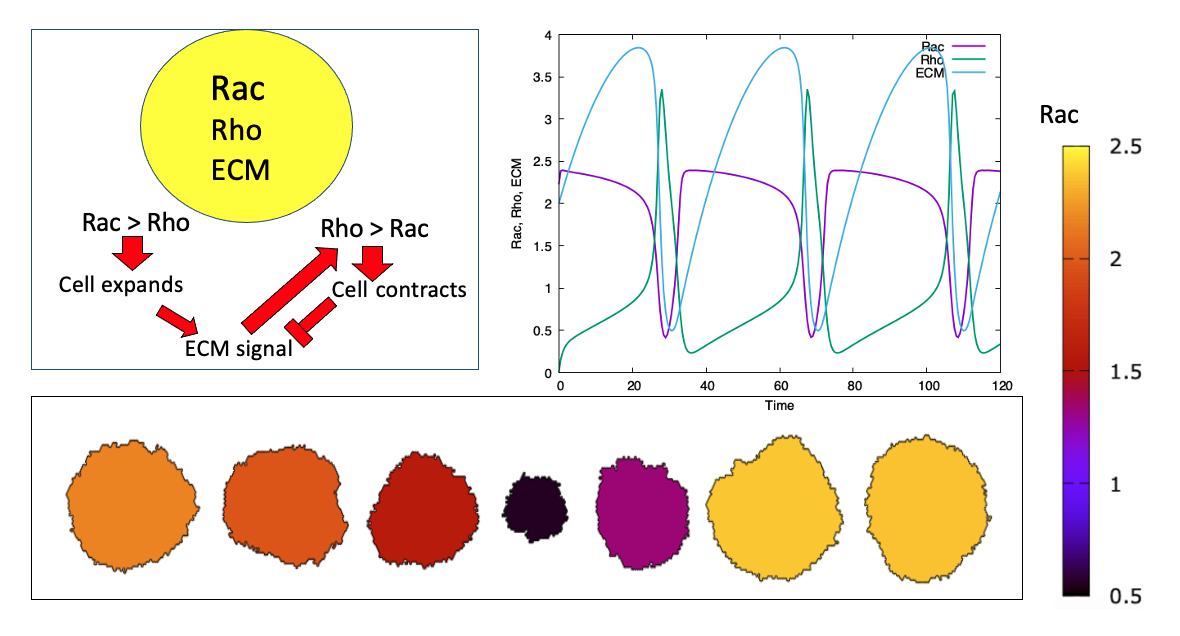}
    \caption{The target volume of a CPM cell is linked to its internal Rac-Rho GTPase levels for a well-mixed Model I Rac-Rho-ECM signaling. The cell contracts and expands accordingly in this oscillatory regime. Top left: schematic diagram. Top right: the time dependent variables in the cell. Bottom row, left to right, the level of Rac (on the color scale at right) in cell shape snapshots at $t=20, 24, 26, 28,  32, 34, 46$. Produced with the open source software Morpheus \cite{Starruss2014} using the 
    file OneLamelShape.xml.  
}
    \label{fig:OneLamelCellSize}
\end{figure}

The well-mixed model is a coarse approximation of the cell signaling, but we can visualize its basic implications by linking the GTPase state to a cell's ``contact area''. To do so, we simulated the well mixed Model I, Eqs.~\eqref{eq:LEK_ModelI}, in a cellular Potts Morpheus simulation, and assumed that the cell target area is high when Rac dominates and low when Rho dominates (Schematic, top left). A representative example of results is shown in panels of Figure~\ref{fig:OneLamelCellSize}. We see the relaxation oscillations typical of such systems (top right panel), and the fluctuations in cell contact area as well as Rac activity level over once cycle in the time sequence of shapes. (Rho activity peaks whenever Rac activity drops.) 

In the paper \cite{JSpark2017,Holmes2017}, the authors considered a well-mixed model with two lamellipods, rather than a single compartment as we show in Figure~\ref{fig:OneLamelCellSize}. Each lamellipod had an internal Rac-Rho-ECM model (similar, though not identical to our Model I). The two were coupled by a constant total pool of Rac and of Rho, and additional terms for competition of the lamellipod size were added. This model could then show regimes of persistent polarity, oscillations, as well as oscillations. A similar example is included in the Appendix Figure~\ref{fig:TwoLamelMorpheusModelResults}. Oscillations and Hopf bifurcations are also possible in the well-mixed variants of Models II and III.

%%%%%%%%%%%%%%%%%%%%%
\section{Setup for the spatially distributed models}

We next consider the full spatially distributed (PDE) models. 
Now there are five PDEs in each of the model variants, to include the PDEs for inactive Rac and Rho. These cannot be eliminated as they diffuse at rates different from the active forms. The full equations are presented in the Appendix for each case.

%%%%%%%%%%%%
\subsection{Spatial geometries}
We consider three distinct geometries, as shown in the schematic, Figure~\ref{fig:SchematicOfModels}. (a) A 1D domain, as in Figure~\ref{fig:SchematicOfModels}b. This is interpreted as a transect across the diameter of the cell, with ends at opposite cell edges. This 1D model can also be interpreted as the geometry of a cell that is confined to a narrow channel as in \cite{Lin-12}or moving along thin ``ECM'' fibers or nano-wires, as in \cite{sharma2017,Padhi2020,Padhi2021}.  
(b) A static 2D circular domain, as illustrated in Figure~\ref{fig:SchematicOfModels}c.  This represents a cell that is ``frozen'' by an inhibitor of the cytoskeleton, or stuck to an adhesive island. In this case, the signaling can take place, but there is no net motion or change of shape. (c) A fully deforming 2D cell, as shown in Figure~\ref{fig:SchematicOfModels}d. In all these cases, we assume Neumann boundary conditions, since no proteins leak out of the cell edges.

%%%%%%%%%%%%%%%%%%
\subsection{Kymographs}
In each case, and for each of the model variants I, II, III, we represent results by a kymograph, as shown in Figure~\ref{fig:SchematicOfModels}e, with time increasing up the vertical axis, and position along the horizontal axis. In 1D, the position is distance along the cell diameter, $0<x<L$. For the 2D cells, the model is solved on a 2D domain but results are summarized by similar kymographs of activity along the cell edge. The``position'' is taken along the (normalized) circumference of the shape, so that $-\pi <x< \pi$ around the cell perimeter. 

Excitable systems that are spatially distributed are known to sustain standing waves, target waves and spiral waves (2D) as well as waves that reflect from domain boundaries. A single standing wave that has high Rac at one domain boundary and high Rho at the opposite end corresponds to a polarized cell. Spiral waves can be recognized by the rotation of the high Rac activity along the domain edge, which in our kymographs, appears as a series of bands where the slope of the band $\Delta x/\Delta t$ is the wave speed along the boundary. Waves that reflect from one edge to the opposite edge show up as a set of peaks and troughs. In all cases, we display the Rac activity on a color scale from low (black or purple) to high (bright orange and yellow). Rho activity (not shown) is the reciprocal; high Rac implies low Rho and vice versa. See Figure~\ref{fig:timeseries-racrhoecm} for an example where all variables are displayed.

\subsection{Parameter sweeps}

%In experimental work on melanoma cells,
In correspondence with Park et al \cite{JSpark2017}, we explore how model behaviour changes when
$\gamma_E$ (rate of ECM-feedback to Rho activation) and $b_R$ (basal rate of Rac activation) are varied. We also vary the constant $K$ in Model I (basal rate of ECM signaling increase), and the corresponding constants $E_t$ in Models II and III
(the maximal adhesion size).

Our results, shown as arrays of kymographs for each model, can be compared to the 2-parameter bifurcation diagrams in Figure 5d in \cite{Holmes2017} (for ``Hybrid Model 3'' in that paper). 
% We show parameter sweeps for each of the three models in what follows.

%%%%%%%%%%%%%%%%%%%%%%%%%
\section{Results: 1D spatial domain}

\subsection{Model I in 1D}

\begin{figure}
    \centering
    \includegraphics[width=\textwidth]{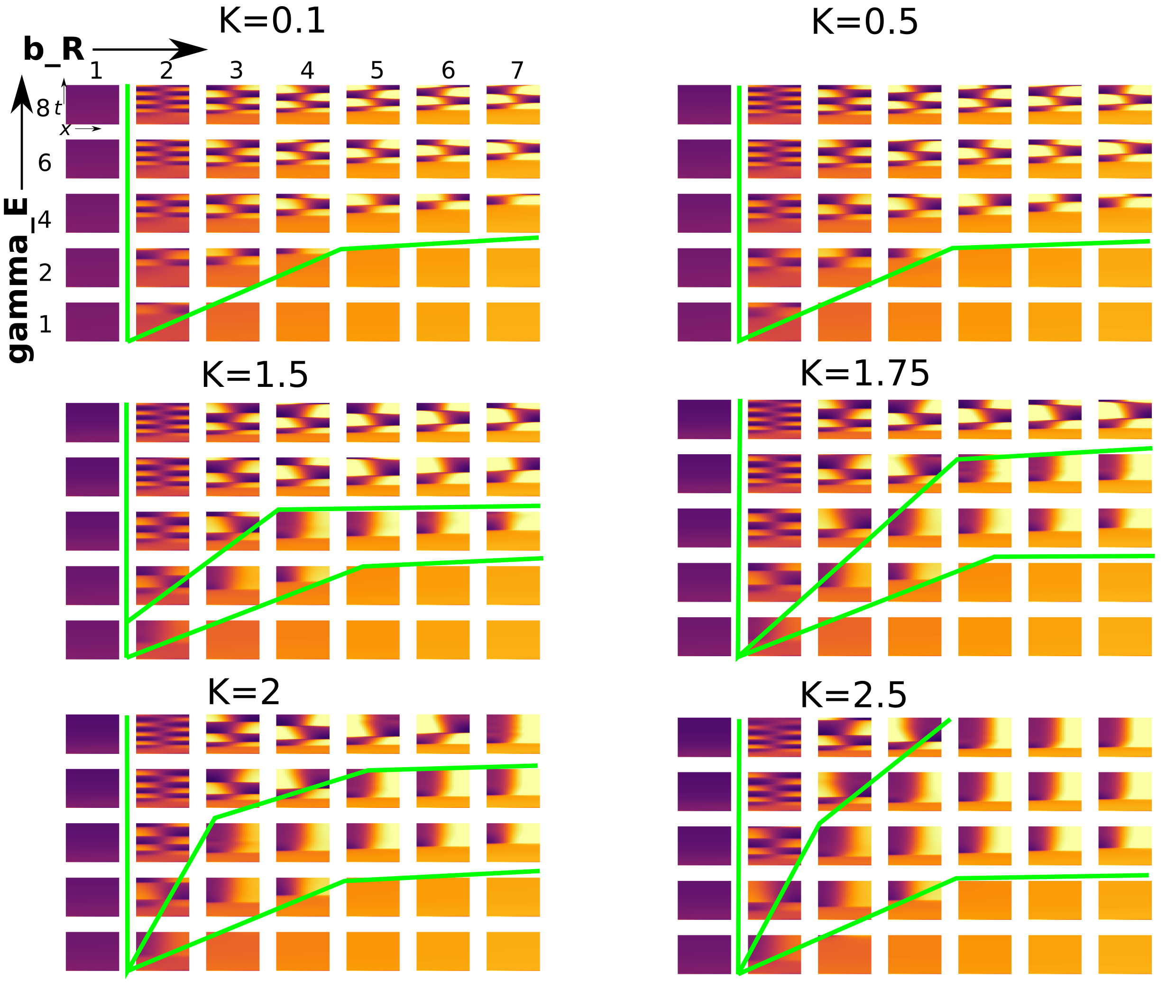}
    \caption{Model I regimes: Behaviour of Rac for the full 1D spatially distributed version of Model I, Eqs.~\eqref{eq:LEK_PDEs1D}, showing parameter sweeps for the basal Rac activation rate $b_R$ and the ECM-induced Rho activation rate $\gamma_E$. Kymographs oriented as in Fig~\ref{fig:SchematicOfModels}e, with $0\le x \le 3$ (horizontally) 
    %The kymographs, are each for a domain of size $L=3$ (horizontal axes) 
    and $0<t<1000$ (vertically upwards). The parameters were varied as follows: $1\le b_R\le 7$ (in steps of 1.0), $\gamma_E=1,2,4,6,8$, $K=0.1, 0.5, 1.5, 1.75, 2, 2.5$. Other
    parameter values: $R_T=3.0,\rho_T=6.0,\delta=1.0, k_E=2 ,\epsilon=0.001, E_0=1.5, n=3, \gamma_\rho=10,\gamma_R=5,k_\rho=0.45,\rho_0=2.4,R_0=1.0$. Heatmap is purple for low and yellow for high Rac activity levels. Approximate regime borders (green) added manually. Produced with Morpheus file Model1RacRhoECMPDEsIn1D.xml}
    \label{fig:ModelI_K-gammeE-bR}
\end{figure}

The spatially distributed Model I is given by the system \eqref{eq:LEK_PDEs1D}.
Results shown in Figure~\ref{fig:ModelI_K-gammeE-bR}, consist of multiple kymographs of $R(x,t)$ (with the same color and axes convention.)
These results agree, broadly speaking, with previous work (``Hybrid'' model in \cite{JSpark2017} Fig 2c in and \cite{Holmes2017} Fig 5d, Model 3). For a fixed value of $K$, we find regimes of low Rac (along the low $b_R$, vertical region of the parameter plane), of high Rac (along the low $\gamma_E$, horizontal region of the parameter plane). Wedged between these are regimes of front-back oscillation and/or persistent polarization. Increasing $\gamma_E$ at intermediate values of $b_R$ results in higher frequencies of oscillation. As $K$ increases, greater swaths to the parameter plane correspond to polarized states, and the wedge of the oscillatory regime shrinks.

\subsection{Model II: Slipbond model in 1D}
\begin{figure}
    \centering
    \includegraphics[width=\textwidth]{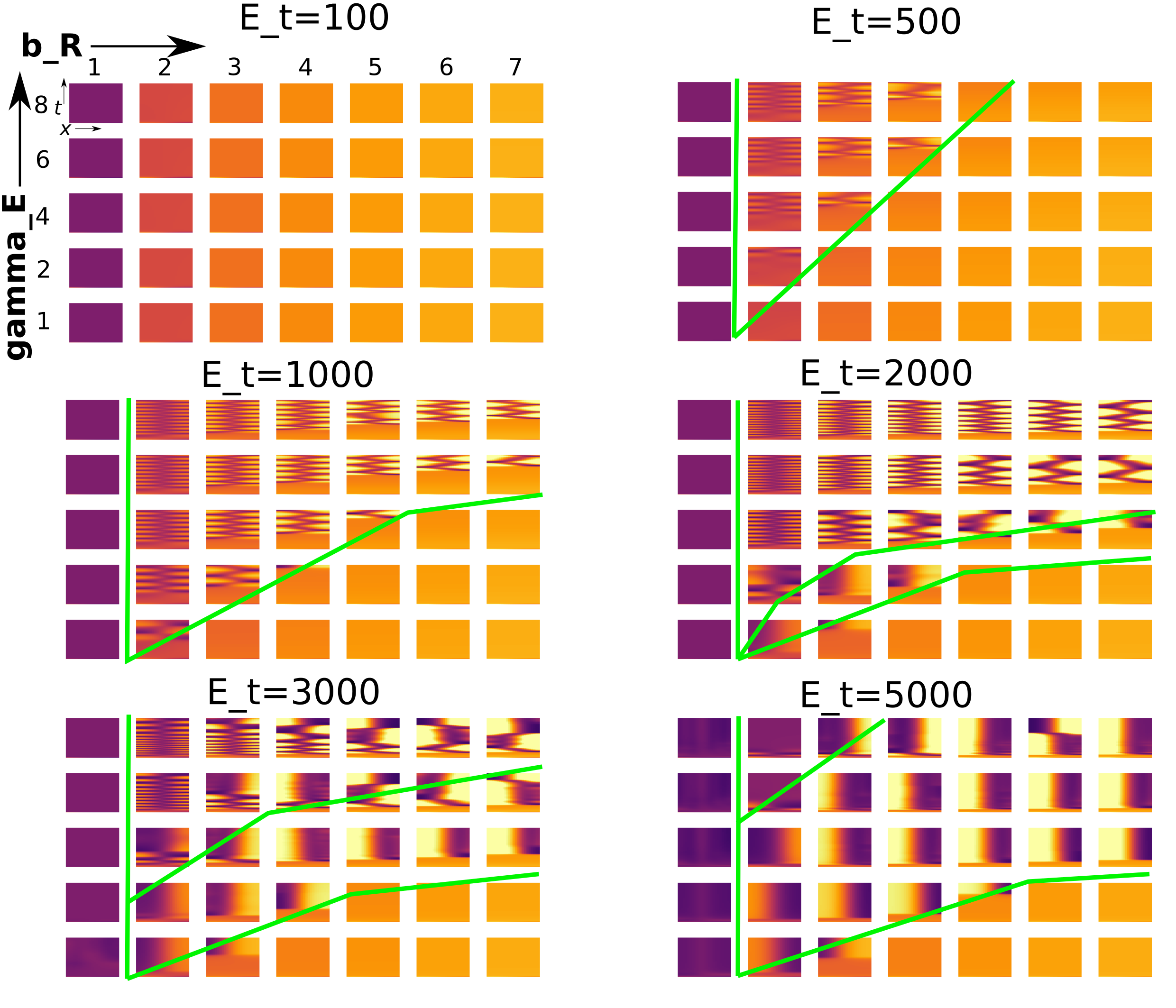}
    \caption{Model II regimes: As in Fig~\ref{fig:ModelI_K-gammeE-bR} but for the slip-bond integrin model (Model II). The parameters were varied as follows: $1\le b_R\le 7$ (in steps of 1.0), 
$\gamma_E=1,2,4,6,8$, max adhesion size,  $E_t$: varied from $100,500,1000,2000,3000,5000$. 
Other parameter values: 
$R_T=3.0,\rho_T=6.0,\delta=1.0, k_E=2 ,\epsilon=0.001, E_0=300, n=3, \rho_0=2.4,R_0=1.0, \gamma_E= 4, K=10,  E_T= 1000, k_0=5, E_s=100, p_s=0.35, \beta_R=1200, \beta_\rho=1600$. Produced by Morpheus xml file Model2SlipBondRacRhoECMPDEsIn1D.xml
 }
    \label{fig:ModelII_Et-gammeE-bR}
\end{figure}

We carried out a similar parameter sweep for the spatially distributed Model II,
Eqs.~\eqref{eq:LEK_PDEs1D} but with ECM as in \eqref{eq:ECM_ModelII_IIIApp} and $d$ given by \eqref{eq:db_slip}
%consisting of the same four Rac-Rho PDEs coupled to the PDE version of the equation for $E$, 
%Eqs.~\eqref{eq:LisannesModelPDEs} 
for slip-bond adhesion sites. Results are shown in Figure~\ref{fig:ModelII_Et-gammeE-bR}, but with variation in the maximal adhesion size, $E_t$ between panels. If the maximal size of adhesions is too small, $E_t \approx 100$, no interesting dynamics is observed, and the domain remains spatially uniform. For $E_t \ge 500$, we see the onset of oscillations. Higher values of $b_R$ delays the onset of the cycles, and increasing $\gamma_E$ leads to period doublings. We see no polar patterns, as yet, but these emerge once maximal adhesion size is tuned to $E_t\ge 2000$, and are even more evident by $E_t=3000$. Finally, by $E_t=5000$, the oscillations are no longer present in the analogous regimes, and the cell is either homogeneous or polarized. Interestingly, a few kymographs demonstrate a polarity reversal in one or another run.

Overall, the results of Model II share similarities with results of Model I, with a few subtle differences. There are similar ``wedge-shaped'' regimes. The behaviour near the low $b_R$ or low $\gamma_E$ regimes are still uniform. There is a similar tendency to settle into a polar state when the adhesions become more dominant (increasing $K$ in Model I versus increasing $E_t$ for Model II).  %When $E_t$ is very large ($\approx 5000$), the cells are mainly polarized, and no oscillations are seem on the same timescale. (A few reversals of direction of polarity can be noted in the lower right panel.) 
As before, the frequency of cycles increases with $\gamma_E$, and decreases with $b_R$. 

%%%%%%%%%%%%%%%%%%%%%
\subsection{Model III: Catch-Slip bond model in 1D}

\begin{figure}
    \centering
    \includegraphics[width=\textwidth]{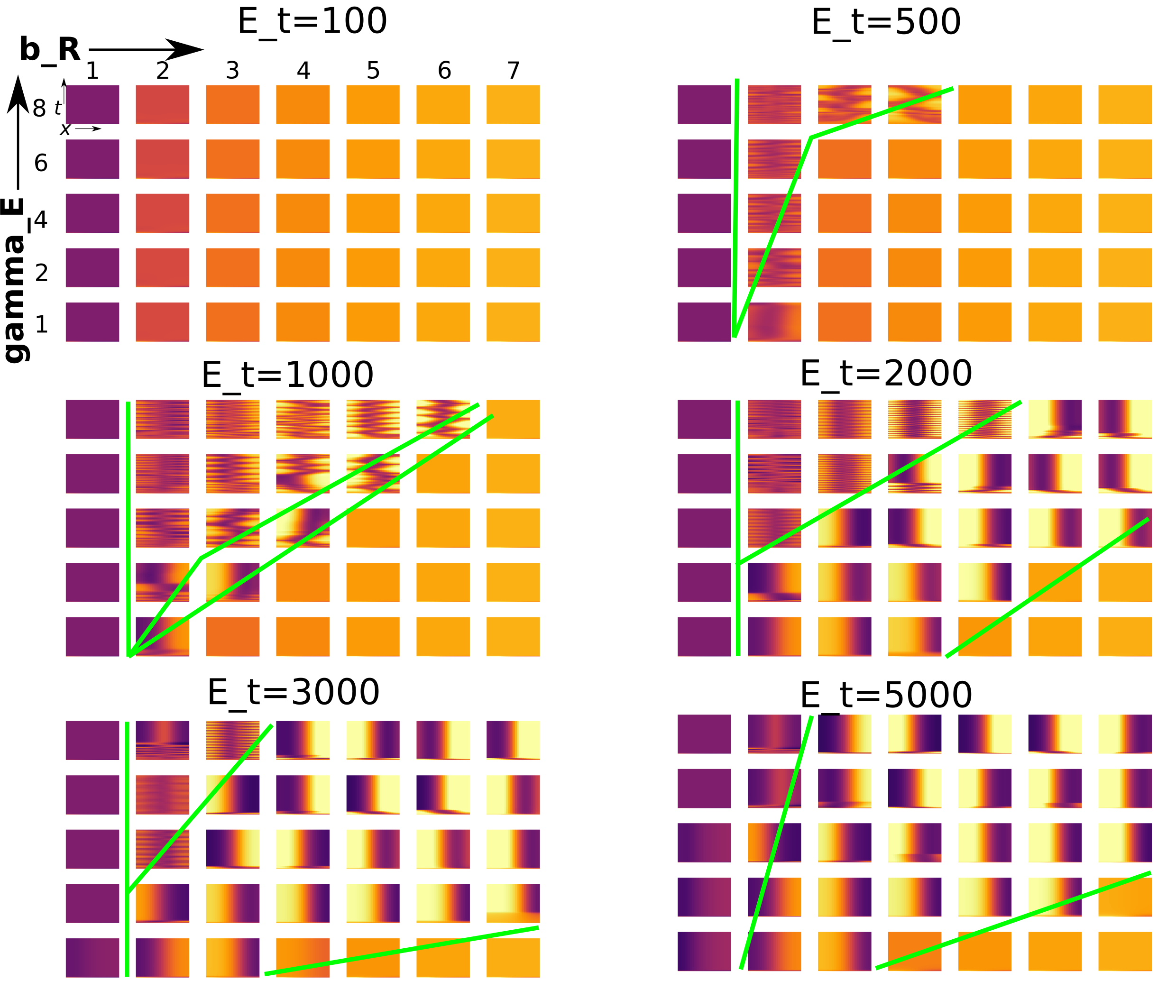}
    \caption{Model III: the catch-slip bond model. As in Figs.~\ref{fig:ModelI_K-gammeE-bR} and \ref{fig:ModelII_Et-gammeE-bR} but for the case where $E$ has the catch-slip bond representation. Similar trends are observed, but for a smaller oscillatory regime. Parameters as in Fig~\ref{fig:ModelII_Et-gammeE-bR} with the following exceptions: $p_s=0.08, \beta_R=1000, k_0=\exp(-7.78), k_{0c}=\exp(4.02)$.
    Produced with Morpheus xml file Model3CatchSlipBondRacRhoECMPDEsIn1D.xml}
    \label{fig:Model3:Et-gammeE-bR-catch}
\end{figure}

Results for Model III are shown in Figure~\ref{fig:Model3:Et-gammeE-bR-catch}. Overall, we observed a substantial decrease in the size of the oscillatory parameter regime. The runs tended to produce polarized patterns of Rac activity, occasionally after a few transient reversals. We found that the oscillatory phenotype only emerges when the back of the cell is in the slip-regime of force, meaning that all adhesion bonds are broken. Only then it is possible for Rac to be sufficiently activated at the back for the front to switch.

\subsection{Significance} Conclusions from the spatial 1D results are as follows. (1) Such models have overall concurrence with the simplified 2-compartment models in \cite{JSpark2017,Holmes2017}, but allow for finer detail on both the dynamics and the spatial distributions. (2) In each of the cases tested, the ECM variable is responsible for emergence of oscillatory behaviour in an otherwise polarizable, multistable system \cite{Holmes2016}. (3) The details of the ECM mechanism, and the ECM equation affect the breadths of some parameter regimes, but otherwise do not significantly change the overall conclusions. The three models tested all share regimes of uniform, polar, and oscillatory behaviours in similar swaths of parameter space. What is more important is the feedback from Rac to amplifying ECM, from Rho to depressing ECM, and from ECM to activating Rho. 

\begin{figure}
    \centering
    \includegraphics[width=0.8\textwidth]{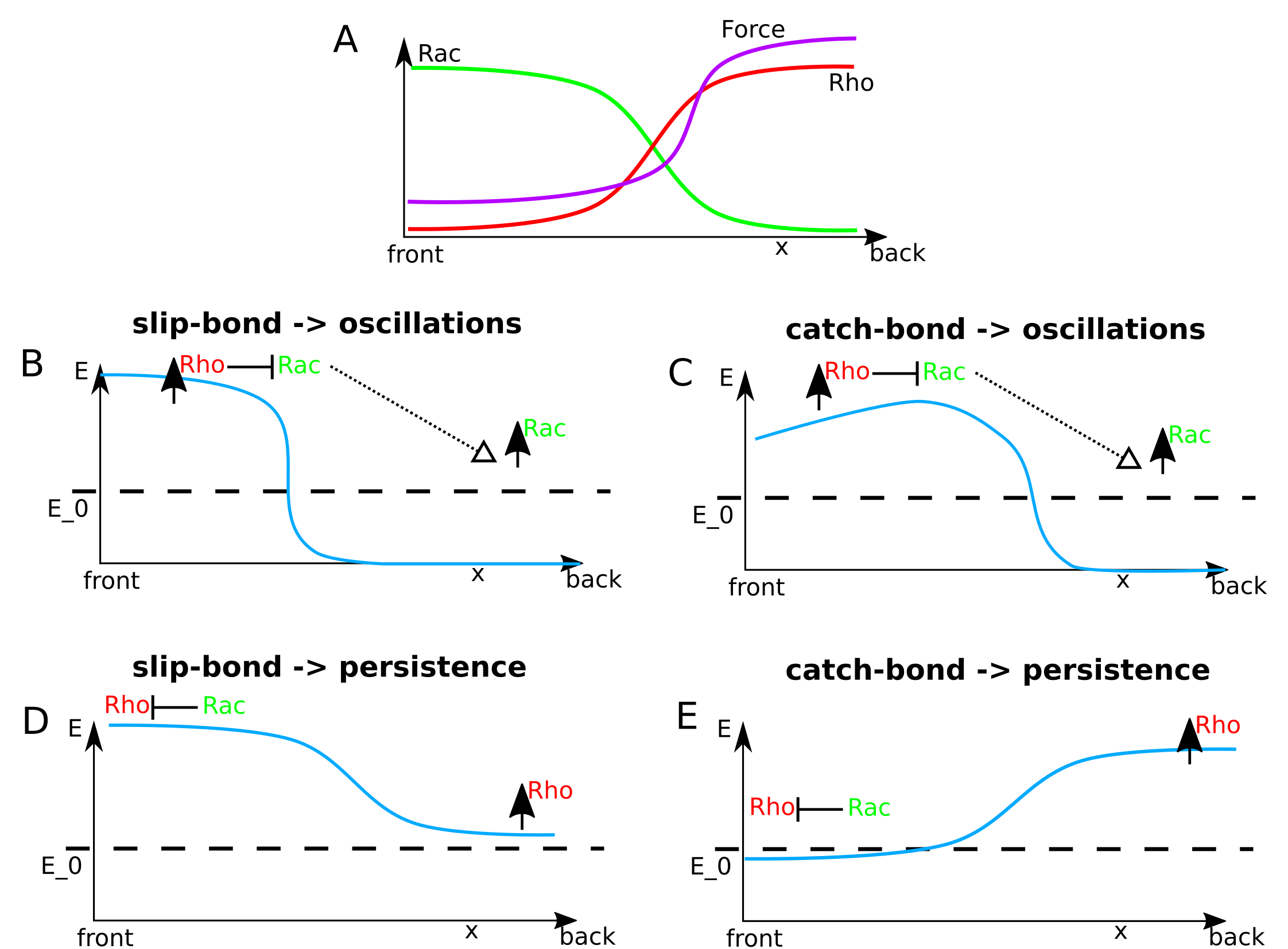}
    \caption{Schematic explanation of how the  slip vs catch-slip integrin dynamics  lead to oscillations or persistence. (A) Typical Rac/Rho and force profiles in a 1D static cell. Highest forces are at the back while low forces are at the front. Increasing or decreasing the total force magnitude leads to high ($>E_0$) vs low ($<E_0$) adhesion, as shown in Figure \ref{fig:steadystates-slip-vs-catch}. (B-C) Adhesion profiles (light blue curve) consistent with oscillations, or (D-E) persistence. (B) Under high force, slip-bonds break at the back of the cell and grow at the front. This upregulates Rho at the front, leading to Rac inhibition. Hence, Rac moves to the back. The pattern repeats, resulting in  front-back oscillation. (C) As in (B), but for the catch-bond model: here adhesion is elevated at some distance away from the front. (D) If the force magnitude is too low, slip-bonds grow at the back of the cell. This reinforces existing Rho activity at the back, causing the polar Rac/Rho profiles to persist. (E) In case of catch-bonds, intermediate force at the back causes maximal adhesion at the back, which as in (D) upregulates Rho and reinforces persistence.  }
    \label{fig:oscillations-explained}
\end{figure}

In Figure~\ref{fig:oscillations-explained} we summarize the mechanisms by which adhesion can cause oscillations or persistence in model II and III. The main idea is that the right level of adhesion is necessary for oscillations to take place. Figure~\ref{fig:oscillations-explained}A shows profiles for Rac, Rho and force $F$ in a 1D polarized cell. Whether the cell polarization persists or exhibits a front-back flip depends on the level of adhesion, which is itself determined by the magnitude of the force and type of adhesion bonds. In general, if the adhesion at the front is higher than $E_0$, it signals to activate Rho. If at the same time, the adhesion at the back is lower than $E_0$, Rho is not sufficiently active to inhibit Rac, allowing Rac activity to invade the rear of the cell, allowing a front-back polarity flip. If however adhesion at the back exceeds $E_0$, Rho is upregulated, which inhibits Rac from moving to the rear, maintaining a persistent cell polarity.

The slip-bond and catch-slip bond models lead to different levels of adhesion (Figure~\ref{fig:oscillations-explained}), but both have regimes consistent with oscillations. Referring to Figure~\ref{fig:steadystates-slip-vs-catch} helps to understand the basis for this behaviour. 
In case of slip-bonds, adhesion is highest at the front and lowest at the back, (Figure~\ref{fig:oscillations-explained}B,D), as adhesion decrease with force (Figure~\ref{fig:steadystates-slip-vs-catch}A). If the force at the back of the cell is above the rupture threshold (Figure~\ref{fig:steadystates-slip-vs-catch}A), oscillations can occur (Figure~\ref{fig:oscillations-explained}B). If the force at the back of the cell is too low, polarity is persistent (Figure~\ref{fig:oscillations-explained}D).

For catch-slip bonds, adhesion is optimal under higher forces  (Figure~\ref{fig:steadystates-slip-vs-catch}B), and hence, maximal at some distance from the cell front, either near the middle  (Figure~\ref{fig:oscillations-explained}C), or at the back (Figure~\ref{fig:oscillations-explained}E). The conditions for oscillations/persistence for the catch-slip bond are similar to the slip-bond, but with different force thresholds (Figure~\ref{fig:steadystates-slip-vs-catch}B).

\clearpage

%%%%%%%%%%%%%%
\section {Results: 2D domains}

From here on, we drop the Original Model I, as it is phenomenological, and concentrate on a comparison of Models II and III. We also restrict attention to the more interesting parts of the parameter regimes.

\subsection{A circular 2D static domain}

\begin{figure}
    \centering
    \includegraphics[width=\textwidth]{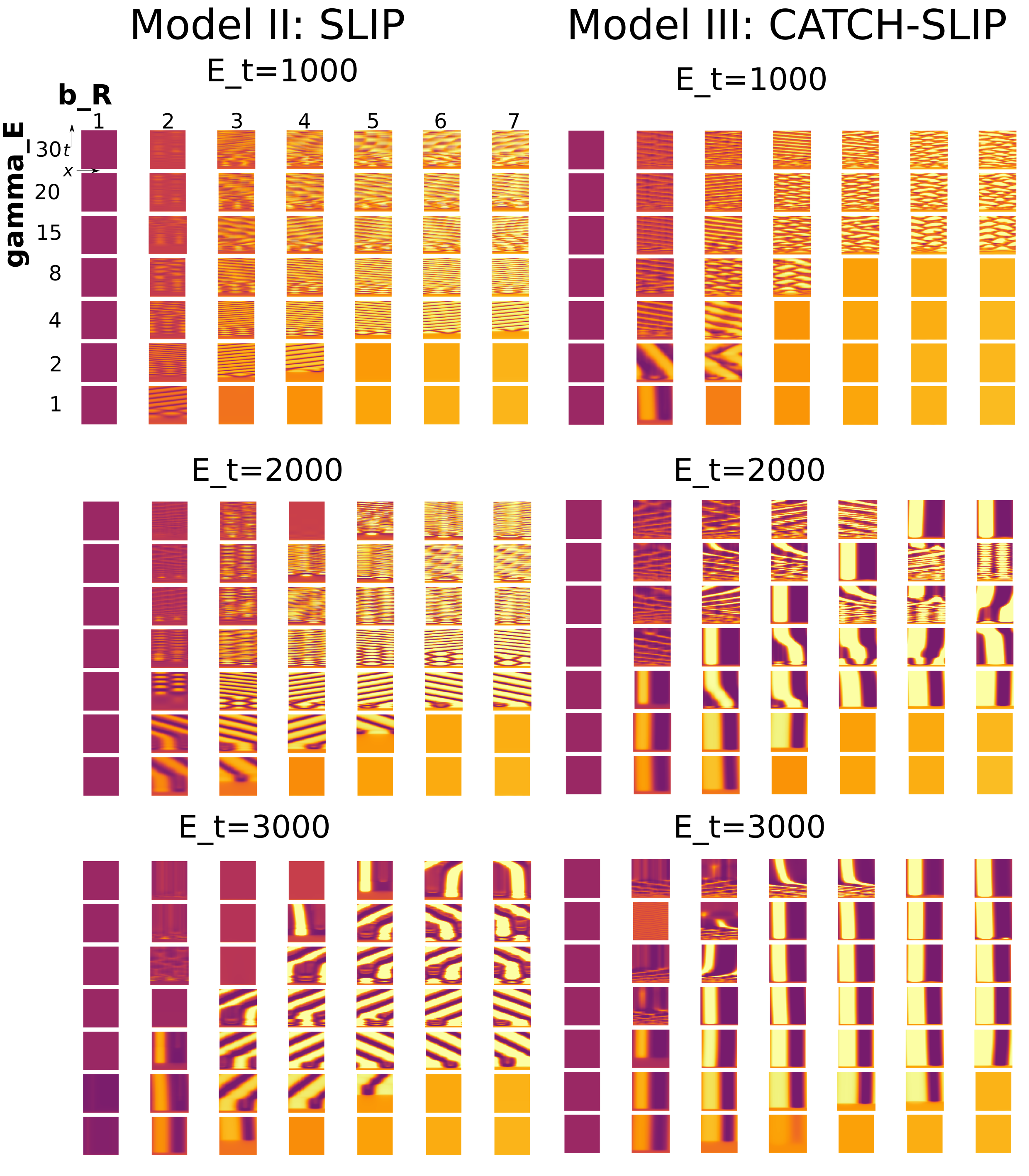}
    \caption{Models II (left) and III (right) in a 2D static circular domain. Kymographs show Rac activity along the rim of the circle. Parameter values and colour scheme as in previous figures. }
    \label{fig:Model2D-v1}
\end{figure}

We simulate the model equations in a fixed circular domain with no flux (Neumann) boundary conditions and initial conditions as described in the Appendix. The kymographs show dynamics along the circular perimeter. The full 2D patterns  consisted of spiral, oscillating or static standing waves. %, and traveling reflecting waves (not shown in 2D).

%, we repeated the parameter sweep of Figures \ref{fig:ModelII_Et-gammeE-bR} and \ref{fig:Model3:Et-gammeE-bR-catch}, 
We carried out parameter sweeps 
for three values of $E_t$ and we added a few larger values for $\gamma_E$. We investigate whether the circular geometry in 2D introduces new phenotypes. 
Results are shown in Figure~\ref{fig:Model2D-v1} for Model II (left column) and Model III (right column) for three values of the adhesion size $E_t$.
%$E_t= 1000$ (top), $E_t= 2000$ (center), $E_t= 1000$ (bottom panel). 

Overall, the behaviour of both Model II and III in 2D shares certain features of the analogous 1D results: regimes of low Rac, high Rac, oscillatory, or persistent polar patterns are evident as before. 
%See Figure \ref{fig:Model2D-v1}. 
The catch-slip model (III) still promotes more polarized persistent behaviour than the slip-bond (II) version, as before.
However, there are several notable differences. (1) for $E_t=1000$, both models, but especially Model III produce a more exotic variety of patterns, including standing oscillations and zig-zag waves, along with random-looking transition patterns. (2)
In place of back and forth 1D cycles, we see traveling waves of Rac activity that get wider as $E_t$ increases. These represent a rotation of Rac along the rim of the circular domain, as would occur when a spiral-wave pattern forms inside. (3) In most cases, back-front oscillatory patterns are transient and evolve into a single spiral around the cell edge. For higher values of $\gamma_E$, the spirals sometimes change direction from counter-clockwise to clockwise, or vice versa. 
Spirals are more likely for higher values of $b_R$, the basal Rac activation and lower values of ECM-Rho activation ($\gamma_E$). Increasing $b_R$ and decreasing $\gamma_E$ both lead to higher Rac and thus wider Rac fronts. 

We find that parameters that resulted in a persistent phenotype in 1D can become spirals in 2D, most notable in the slip-bond model.  
We also find intermediate phenotypes, where spirals are followed by a few transient front-back oscillations, that then evolve into a spiral again. Without carefully observing the pattern, this behavior looked seemingly random at first.

Finally, we note that the boundary between regimes is less sharp in 2D than in 1D results. For example, in Model III for $E_t=2000$, the oscillatory and persistent regimes appear to bleed into one another. This may stem from coexistence of multiple steady states (migratory phenotypes) in a given transition zone in parameter space.
This kind of coexistence has already been discussed in a simpler model of Rac and Rho alone in \cite{Holmes2016} where patterns were far simpler (uniform low, high, or polar).
The 2D geometry appears to accentuate this possibility.

In conclusion, the 2D geometry gives rise to more complex patterns, such as spirals. In a motile cell, spirals could be indicative of circular motion. Or, cell shape changes may transform spiraling into front-back oscillations or stabilize the spiral into a persistent front. Consequently, we asked how the patterning changes as we allow the 2D domain to deform.

%%%%%%%%%%%%%%%%%%%

\subsection{A 2D deforming domain}

Models II and III, 
%given by the PDEs \eqref{eq:LisannesModelPDEs}, 
are now solved over a similar range of parameters, but in a 2D deforming domain, representing the top-down view of the cell shape.
Briefly, domain deformation was tracked using a custom-built cellular Potts model (CPM) calculation with an in-built PDE solver, as described in the Appendix. (As of this writing, open-source packagew such as Morpheus \cite{Starruss2014} and CompuCell3D \cite{CompuCell3D_2005} do not yet have a PDE solver for a deforming CPM cell.)

To link GTPase activity to the evolving cell shape, we assumed that high Rac activity at the cell edge promotes local outwards protrusion of the edge (see, e.g. \cite{Zhan2020}), whereas high levels of Rho lead to inwards contraction as in \cite{YueLiu2020}, bypassing the explicit representations of actin, myosin, and other cellular components that were included in previous work \cite{Maree-06}. In distinction with the coarse-grained results of Figure~\ref{fig:OneLamelCellSize} where the entire cell spreads or contracts, here the spatial distributions of the GTPases is resolved in detail. Hence protrusion and contraction of the cell edge are localized to spots on the perimeter where there is high Rac or high Rho, and we can track both cell polarization, change of overall shape, and motility. Details of the custom built CPM computation are provided in the Appendix.

\begin{figure}
    \centering
    \includegraphics[width=\textwidth]{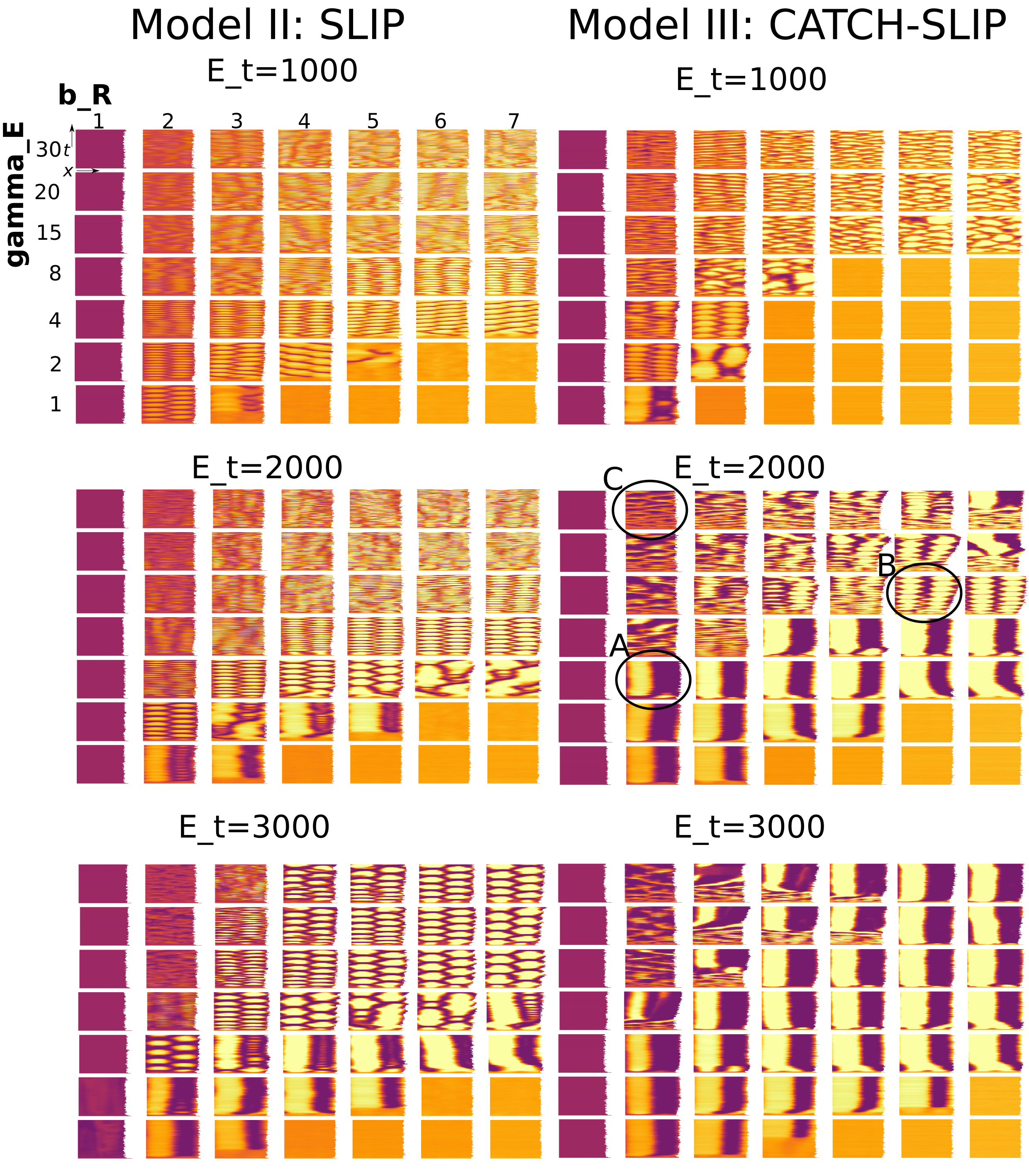}
    \caption{As in Figure~\ref{fig:Model2D-v1} but for a 2D deforming cell. Parameters as in Table \ref{tab:pars}. The same two models and parameter sweeps are shown. The shapes of three sample cells (circled above) are shown in Fig.~\ref{fig:threephenotypes-v1}. Movies can be found here \url{https://imgur.com/a/Btw4z9H}}
    \label{fig:cellmoves-Model2D-v1}
\end{figure}

\begin{figure}
    \centering
    \includegraphics[width=0.6\textwidth]
    {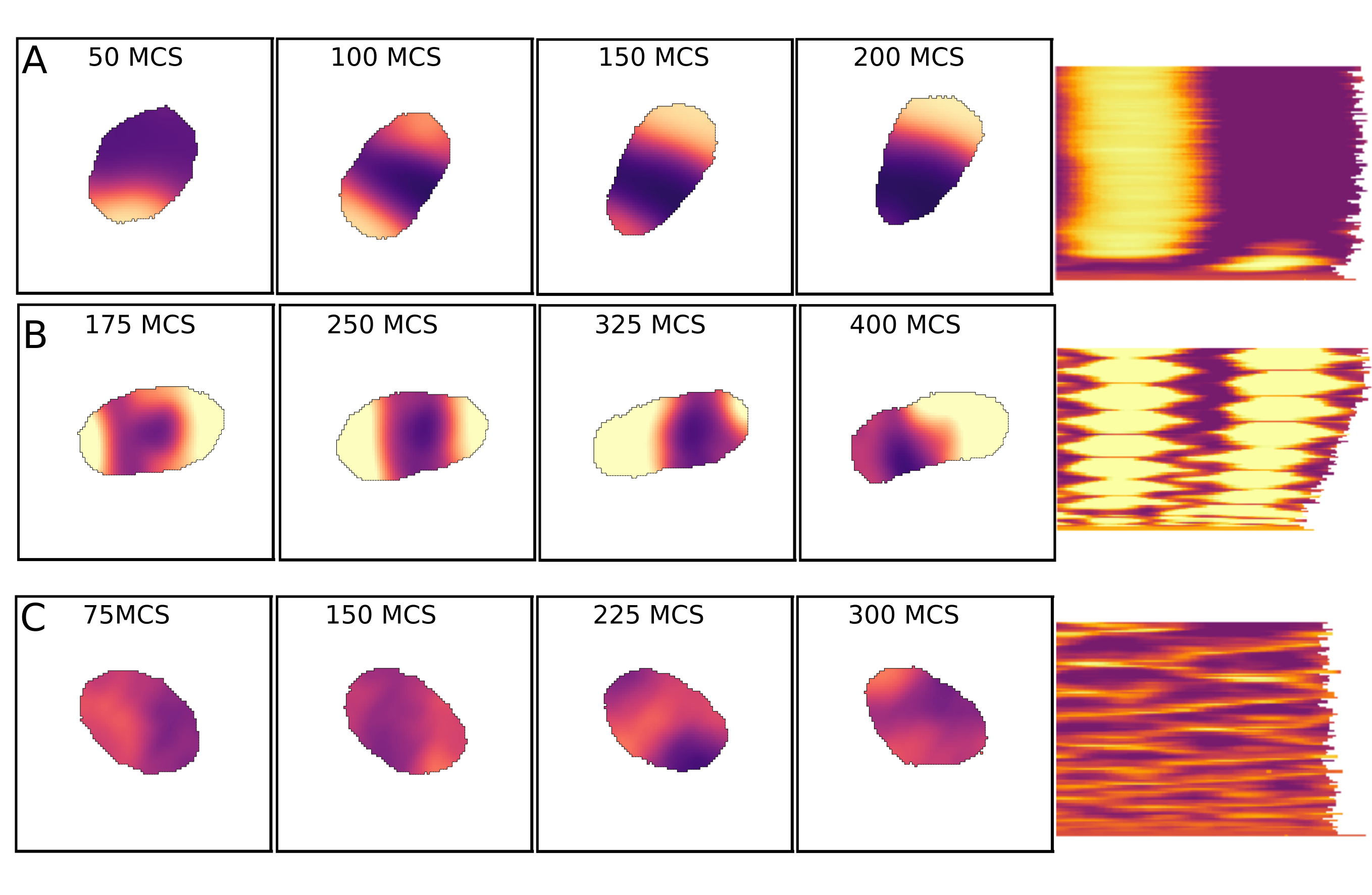}
    \caption{Sample cell shapes and Rac activity in a brief time sequence.
    The three parameter settings correspond to circles in Fig \ref{fig:cellmoves-Model2D-v1}. (A) Persistent cell. (B) Oscillatory cell. (C) Random cell protrusions. The kymographs (right) show the long-time behaviour for each case, and are magnified views of the circled images in Fig \ref{fig:cellmoves-Model2D-v1}. Note that the expanding perimeter of the oscillatory cell is seen in the kymograph for (B). Movies can be found here \url{https://imgur.com/a/Btw4z9H}}
    \label{fig:threephenotypes-v1}
\end{figure}

Using the same parameter sweeps as before for Models II and III, we show results in Fig~\ref{fig:cellmoves-Model2D-v1}, with selected sample cell-shape time sequences in Fig.~\ref{fig:threephenotypes-v1}.
As before, kymographs track the Rac activity level along the cell edge (dotted white curve in Figure~\ref{fig:SchematicOfModels}d). Because the domain deforms, and its perimeter length fluctuates slightly with time, the right edge of each kymograph in panels of Fig~\ref{fig:cellmoves-Model2D-v1} appear slightly ragged.

By comparing patterns in the 2D deforming domain (Figure~\ref{fig:cellmoves-Model2D-v1}) to the identical conditions in the 2D 
static domain (Fig.~\ref{fig:Model2D-v1}), we can identify the effect that domain dynamics has on internal patterns. 
Regimes of spiral waves in a static domain tend to become front-to-back oscillations in the deforming domain.
This is evident in the Slip-bond model (II) for $E_t=2000, 3000$ where the slanted yellow-purple bands that represent spiral waves in the static case (Fig.~\ref{fig:Model2D-v1}) are replaced by the ``honeycomb'' kymograph patterns that depict front-back oscillations (Fig.~\ref{fig:cellmoves-Model2D-v1}).
For the slip-bond with $E_t=3000$, a larger regime of polarization emerges in band at low $\gamma_E$ that was previously only spiral waves. The cell deformation appears to dampen spirals in favour of either front-to-back cycles or stable polarity.

The slip-bond model produces much more regular oscillations than the catch-bond model. It is easy to note that the frequency of oscillation is reduced at higher ECM $E_t$ (compare $E_t=2000$ with $E_t=3000$). On the other hand, the ECM-Rho feedback strength ($\gamma_E$) increases the frequency. So, the feedback strength and ECM binding rate, both of which regulate the amount of signaling, have opposite effects on the oscillation frequency. The explanation is that for higher values of $E_t$, the cells need to apply more force to break the bonds at the back of the cell to allow a front-back switch to occur. Higher values of $\gamma_E$ makes the cell more sensitive to weak signals from the ECM, allowing for quick oscillations.

For the catch-slip bond model (III), at $E_t=1000$, results in static and deforming cell domains are somewhat similar, except for the corner at low values of both $b_R$ and $\gamma_E$, where spirals (in the static domain) turn into cycles (deforming domain). Polar patterns still tend to dominate at $E_t=2000, 3000$, but we find an increased swath of irregular and random oscillations for $E_t=2000$ when the cell can deform.  

We selected three examples of interesting dynamics (indicated by circles in Fig.~\ref{fig:cellmoves-Model2D-v1}) to track in the full 2D simulations. These include (A) a persistent polar case, (B) front-back oscillations, and (C) random activity pattern. The cell shape dynamics are shown for a short time sequence in Fig.~\ref{fig:threephenotypes-v1}, together with a more detailed view of the corresponding (long-term) kymographs, to demonstrate the overall outcomes. In (A) we see a cell that polarizes, and starts to migrate directionally. The polarity persists over time. For (B), high Rac activity oscillates between two cell ends. This causes the cell to elongate. Hence, the perimeter of the cell also increases with time, as seen in the gradually expanding envelope of the kymograph. This cell fails to have significant net migration. 
In (C), the zone of active Rac continues to move around the cell irregularly, so protrusion/retraction is undirected, and the cell fails to migrate. The kymograph demonstrates many interpenetrating waves along the cell edge, without clear periodicity.

We can understand some of the results of Fig.~\ref{fig:cellmoves-Model2D-v1} from the actual cell shapes. First, for a cell with initial spiral internal dynamics, the Rac-induced edge protrusion can result in slight elongation of the cell. This tend to break symmetry, damping the spirals that arrive at the domain boundary. An elongated domain then favours the  oscillations that dominated the previous 1D results, and the oscillations, in turn make the cell longer and contribute to self-amplification. Cell edge deformation can also push spirals into a single front, entering the persistent regime. This is visible in the slip-bond model with $E_t=3000$, where many patterns were spirals in the static 2D domain, but have become persistent in the deforming domain. 
Most notably the catch-bond model in combination with shape deformations causes irregular oscillations. In some cases, these oscillations ultimately evolve into a single persistent front.

In conclusion, the 2D deformable simulations show that cell motility can change the phenotypical outcome expected from static 2D simulations. This can explain the huge variability in the three phenotypes of the same cell types on the same kind of ECM, something observed also in \cite{JSpark2017}.

\section{Results: effect of cell stiffness on patterning dynamics}

Finally, we asked how the biophysics of the cell and the properties of the domain affect the internal patterning of the signaling activity. To do so, 
we reran parameter sweeps for variants of the CPM simulations with distinct CPM parameters. (1) We decreased the feedback from Rac/Rho to cell edge protrusion/retraction. This manipulation would be related to the responsiveness of actin polymerization at the front (activation of WASp and availability of Arp2/3 downstream of Rac) or the downstream recruitment of ROCK and myosin activation downstream of Rho).
(2) We also changed the stiffness of the cell. For this purpose, we modified the relative values of the CPM area and perimeter constraint and cell-medium interfacial energy \cite{Magno2015}.
(3) Some cell types have far more rapid signaling responses than others. For example, GTPase responses in neutrophils occur on a timescale of seconds, whereas in fibroblasts, HeLa cells, and melanoma, the timescale is far longer (many minutes, hours). To investigate this, we  
reduced the number of PDE iterations per MCS by a factor of 2. This is equivalent to slowing down the signaling kinetics relative to cell motion.

\begin{figure}
    \centering
    \includegraphics[width=\textwidth]{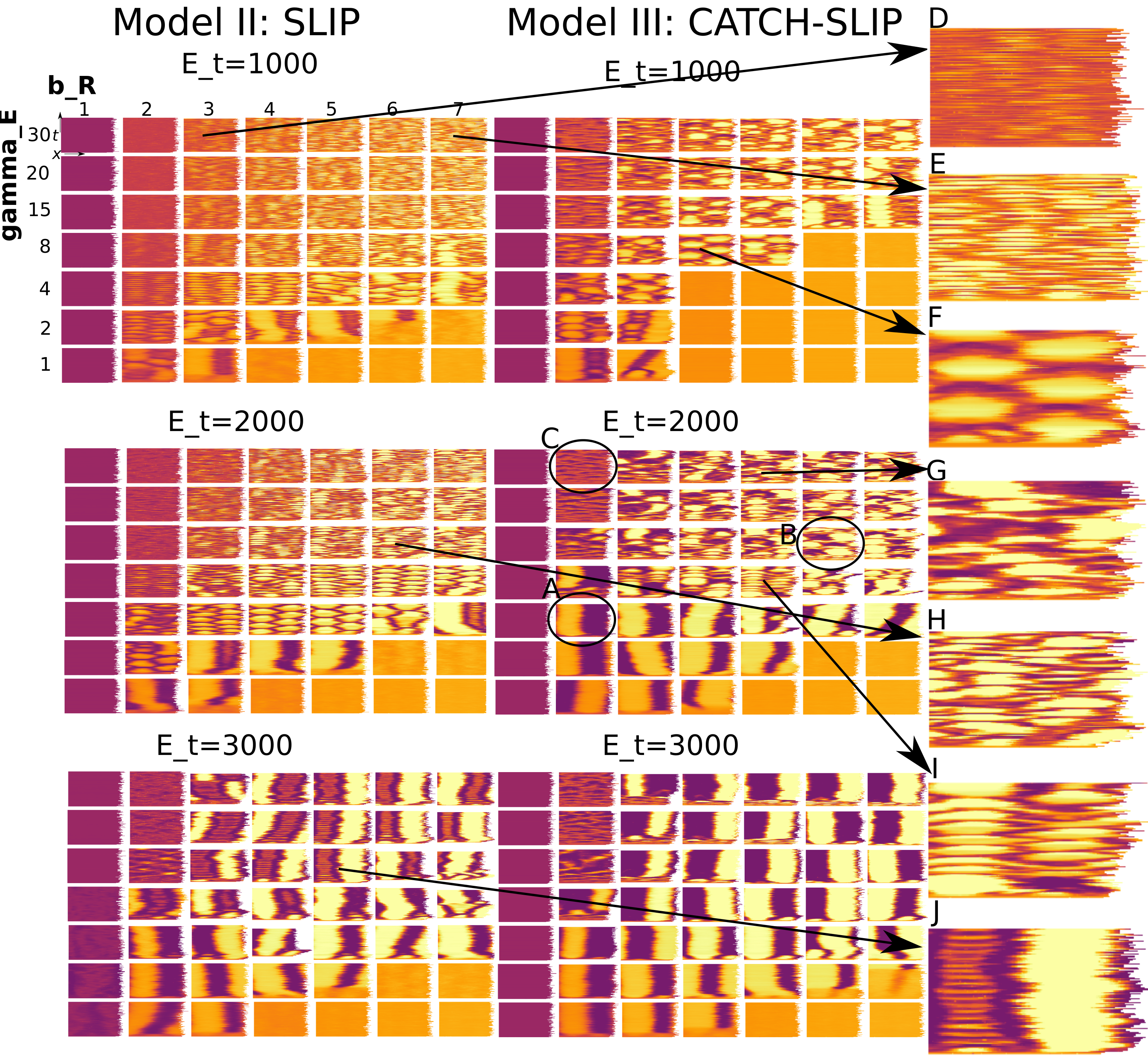}
    \caption{Rac signaling in 2D deforming cell with distinct biophysical properties. Parameters as in Table \ref{tab:pars}, except with $n_i=400,N=2000,\lambda_A=1.5,\lambda_\epsilon=10,P_0=400,\lambda_P=3,J=20000,\beta_G=0.25$. Movies can be found here \url{https://imgur.com/a/Btw4z9H} }
    \label{fig:cellmoves-Model2D-v2}
\end{figure}

\begin{figure}
    \centering
    \includegraphics[width=0.6\textwidth]{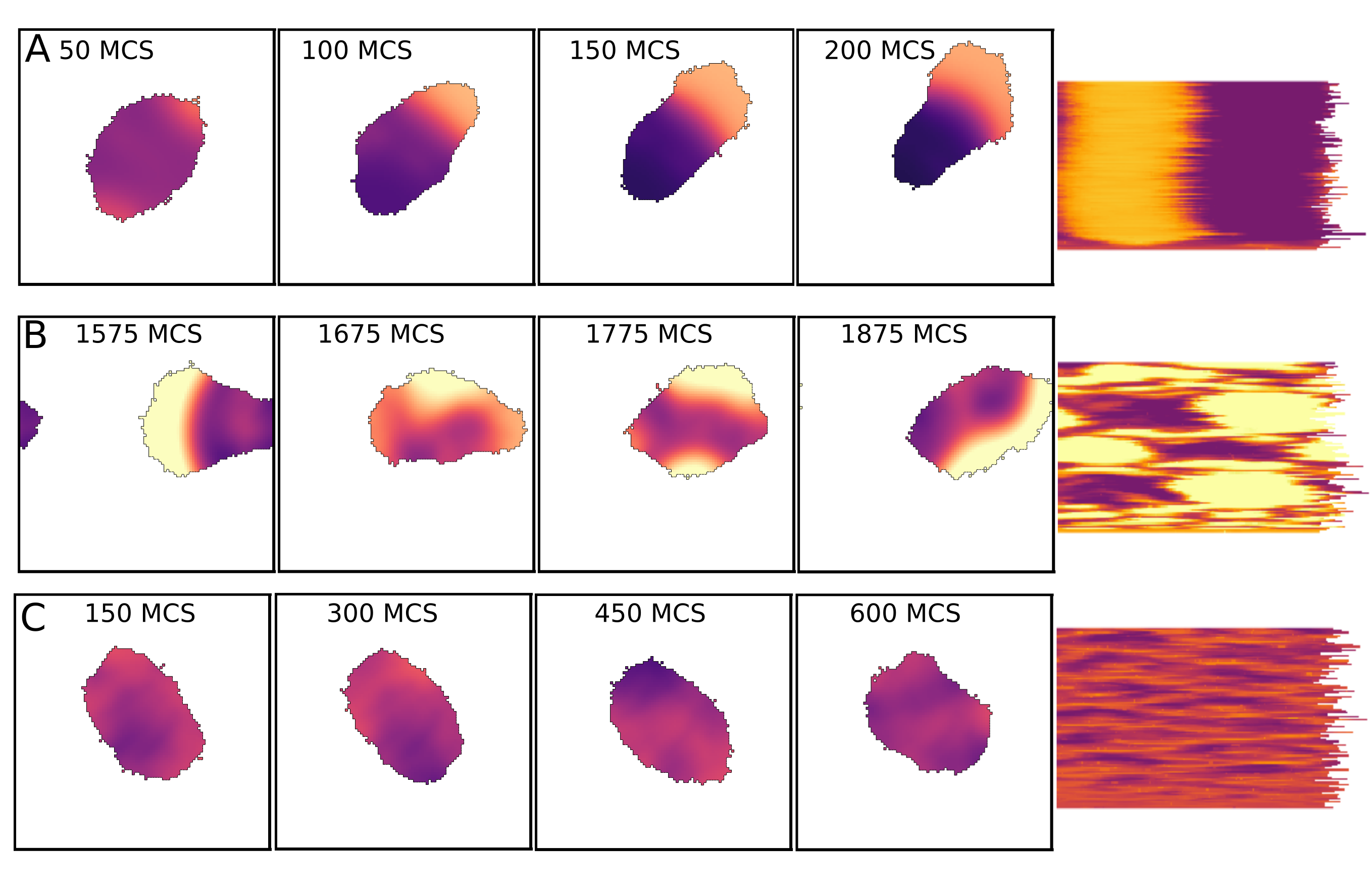}
    \caption{Three phenotypes from the circled examples in Fig \ref{fig:cellmoves-Model2D-v2}. Movies can be found here \url{https://imgur.com/a/Btw4z9H}}
    \label{fig:threephenotypes-v2}
\end{figure}

Results of these tests, shown in Figure \ref{fig:cellmoves-Model2D-v2} demonstrate  that biophysical parameters such as cell stiffness greatly affect internal reaction-diffusion patterning. We highlight a few of these simulations by small panels labelled (A) to (J) in Figure \ref{fig:cellmoves-Model2D-v2}. We demonstrate the full cell-shape dynamics of several typical cases in Figure~\ref{fig:threephenotypes-v2} (See also video links in the caption of Figure~ \ref{fig:cellmoves-Model2D-v2}.) 

As before, we observe persistence (A) and very regular oscillations (F). But, we now also observe patterning in what were formerly uniform Rac parameter regimes. In addition, we find various interesting intermediate phenotypes. For instance, we observe a cell that has irregular oscillations (B,G,H,I), where the period of the front-back oscillations vary. Some cells exhibit a kind of stick-slip motion (J), where Rac transiently appears at the back of the cell and briefly halts the forward motion, but the cell remains overall persistent. We also observe  phenotypes that oscillate and spiral simultaneously or consecutively (D,E). In these, transient spirals and oscillations briefly push the cell edge outwards.

\begin{figure}
    \centering
    \includegraphics[width=0.6\textwidth]
    {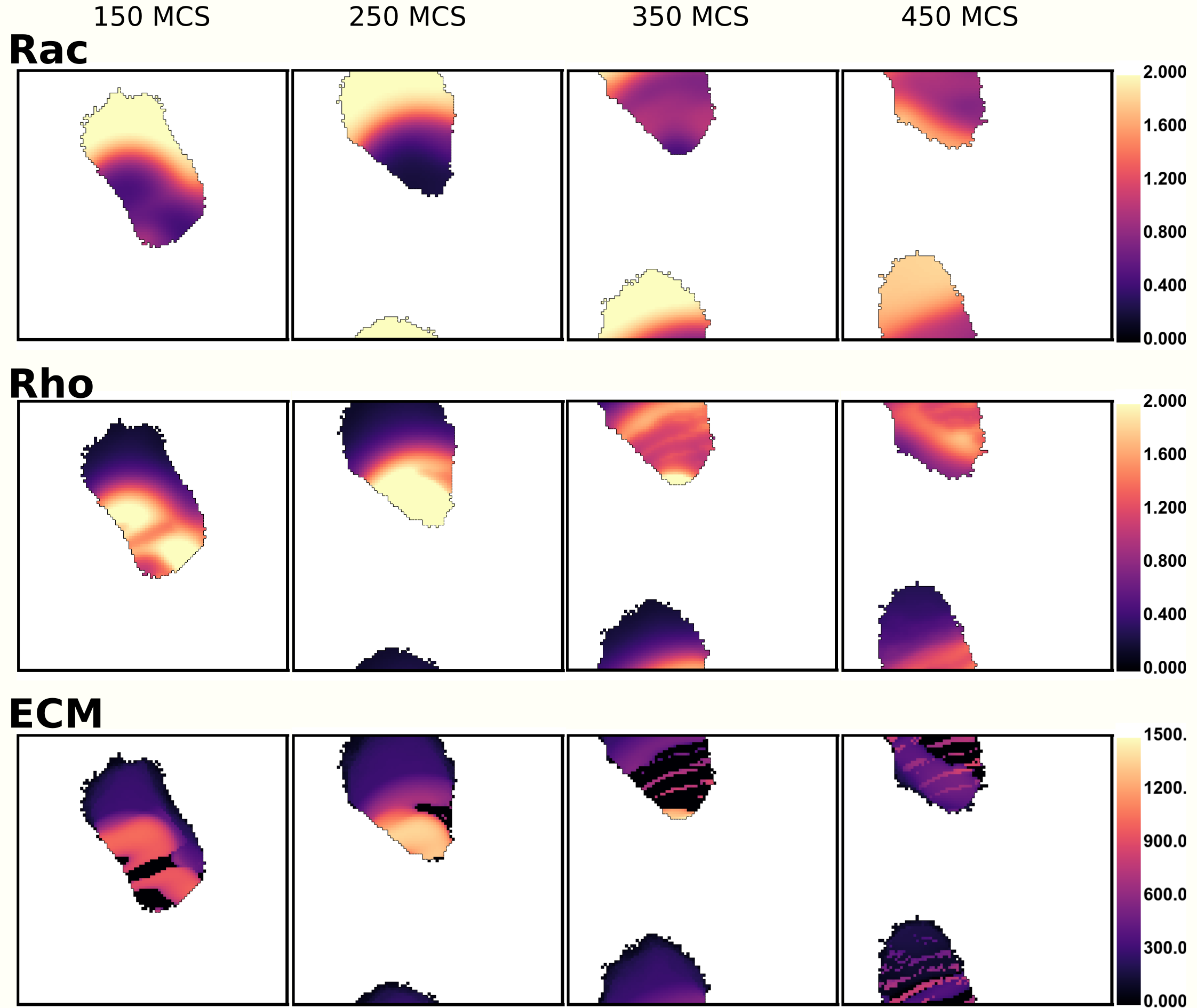}
    \caption{Rac, Rho signaling and ECM adhesion during cell motility of example (I) from Figure \ref{fig:cellmoves-Model2D-v2}.  Movie can be found here \url{https://imgur.com/a/p8zxH64}}
    \label{fig:timeseries-racrhoecm}
\end{figure}

Among the phenotypes, we see varying degrees of periodicity and spatial regularity. For example, in Figure~\ref{fig:timeseries-racrhoecm}, we show a time series of a persistently moving cell (labeled (I) in Figure \ref{fig:cellmoves-Model2D-v2}) that suddenly flips its front. We show some time-steps right before the front-back switch and visualize Rac, Rho and the ECM field. As explained in Figure \ref{fig:oscillations-explained}, a front-back switch can occur provided adhesion is sufficiently high at the front and, simultaneously, sufficiently low at the back. This is seen in the 2D moving cell in Figure \ref{fig:cellmoves-Model2D-v2} and its accompanying movie. As the cell moves, it creates adhesions at the front (150 MCS). At the back of the cell, forces are high due to the presence of Rho. Due to the catch-bond dynamics, these forces stabilize the adhesion at the back (250 MCS). However, at some point, the forces breach the threshold that breaks adhesions at the back (350 MCS). Then, the ECM-induced Rho activation at the front results in repolarization (450 MCS). Interestingly, some mature adhesions are very long-lived, even at the back of the cell ($t\ge$350 MCS). 
Such repolarization takes place periodically,  every time there is sufficient adhesion breakage at the rear of the cell. 
The precise combination of GTPase dynamics and biophysical cell parameters, determines how often and where adhesions tend to break, and thus when and where the Rac front will appear. This results in the various of phenotypes we observe.

\clearpage

%%%%%%%%%%%%%%%%%
%%%%%%%%%%%%%%%

%%%%%%%%%%%%%%%%%%%%%%%
\clearpage
\section{Discussion}

%\LEK{ \textbf{Quick summary of paper}}

In this paper, we formulated a spatially-distributed set of models for signaling and ECM feedback, motivated by the experiment-theory work of \cite{JSpark2017, Holmes2017}.  We added more detail in several ways: (1) We considered a fully spatio-temporal (PDE) model in place of the original 2-compartment models in \cite{JSpark2017, Holmes2017}. (2) We revised a generic ECM-feedback model to one based on the biophysics of integrin adhesion bonds, allowing for more direct experimental validation.
(3) We investigated how geometry in 1D, 2D, and a deforming 2D cell affect the GTPase patterning dynamics. For the deforming cell simulations, we used the Cellular Potts formalism. Previous work of this type includes \cite{Maree-06} (a Cdc42-Rac-Rho model with F-actin barbed ends)  and \cite{YueLiu2020} (a single-GTPase toy model with F-actin feedback).

By explicitly formulating the roles of local cell forces, integrin dynamics, and cell shape changes, we could predict observed phenotypes, new phenotypes, and phenotypical switches. While many details of our models differ from those of \cite{JSpark2017,Holmes2017}, 
we find similar basic regimes of persistent, oscillatory, and random behaviour, together with new intermediate behaviours that were not previously apparent. 
We examined the models in a hierarchy of well-mixed, 1D, 2D, and deforming 2D. This helps to build up some basic understanding before exploring the full complexity of the spatial patterning.

%\LEK{ \textbf{About the pattern dynamics}}

As previously observed, Rac-Rho mutual antagonism results in bistability \cite{Holmes-15}, and feedback from other influences (here the ECM) allows for oscillations and other dynamics in single cells \cite{holmes2012regimes,Holmes2017,Buttenschon2020} or in a cell sheet \cite{Zmurchok2018}.   %How 
The precise nature of the feedback has a limited influence, as we observed in our three model variants. The key idea is that the relative Rac/Rho activity levels determine the magnitude of contact with the ECM, whereas the slow feedback from the ECM signaling leads the system ``around a hysteresis loop'' by gradually tuning the Rac-Rho competition. 
 As described in \cite{Holmes2016,Zmurchok2020}, there are regimes in which several types of steady state patterns coexist (e.g. polar and oscillatory, in our case). This shows that cells can have multiple repertoires of behaviour, even when their parameters are the same. A single ``genotype'', that corresponds to a given set of kinetic rates, can give rise to multiple phenotypes.

 We found that geometry has significant impact on the dynamics of patterns. For example, in 2D domains, we see regular and irregular internal waves such as fronts and spiral waves.
 Spirals have been intensively studied in pattern-forming chemical and biological systems since the 1970s \cite{Winfree1972,Tyson1989}. Analysis of such patterns by various mathematical, geometric, and physics-based methods \cite{Winfree1991,Cross1993,Keener1986} is typically restricted to static domains. It is well-known that bistable reaction-diffusion PDEs with slow negative feedback can give rise to traveling waves \cite{Rinzel1982}, and that addition of a conservation condition can cause those waves to stall \cite{Mori-08}.
Presence of spiral-waves inside cells have been connected to feedback between actin and other components \cite{Vicker2000,Whitelam2009,Bretschneider2009,Bernitt2017,holmes2012regimes}. A survey of traveling waves in actin dynamics more generally can be found in \cite{Allard2013}.

Our most interesting observation is the interplay between a deforming domain and the internal reaction-diffusion dynamics. The regimes of behaviour in static 2D domains shift when the patterns interact with an evolving domain boundary. To some extent, this is expected from previous models of static and deforming 2D cells. For example, static and deforming domains are compared in \cite{Liu2019} for several basic models of a single GTPase coupled to negative feedback from F-actin. Dynamic patterns of a Rho network on a static cell shape were modeled and analyzed mathematically in \cite{Bolado-C-2020}. In \cite{eroume2021}, a circuit of Cdc42, Rac, Rho, and phosphoinositides was simulated on static cell shapes; it was shown that certain shapes can reverse polarization gradients.  In one of the original fully deforming domain simulations \cite{Maree-12}, it was shown that domain deformation can accelerate internal RD dynamics of these intermediates. Neumann boundary conditions used to depict sealed cell edges set constraints on the level curves of internal concentrations: those level curves must be orthogonal to the boundary, so small boundary changes dramatically affect the internal distributions. 
A full mathematical exploration of PDEs in deforming domains is still in its infancy, and our observations provide motivation for further analysis of such patterning dynamics.

We note that to obtain our results, we set the diffusion length of the active forms of GTPase to be 1/10 of those of the inactive form. This assured that the length scale of the Rac-front would be of the right width relative to cell diameter, leading to front-back cycles instead of waves. Decreasing the diffusion length of the active form (or increasing cell size) generates multiple interacting waves. The kymographs of the Rac behavior at the cell edge in the 2D static domains were similar, but the wave patterning in the cell interior changed. In the deforming cell, this leads to even more irregular phenotypes than shown here, as multiple co-existing waves hit the cell edge in different locations.

We found that cell deformation can reinforce patterns that were only weakly stable on a static domain. For example, cells that elongate tend to become more stably polarized;  the Rac front tends to concentrate at one end of the oval, rather than slowly spiraling around.  There is then a positive feedback between polarization and further polarization, favouring persistence. In a different parameter regime, cell elongation can also evolve spirals into front-back oscillations.

 In correspondence with the two-compartment model, we find the persistent and oscillatory phenotype, and cells with either uniform high Rac or uniform low Rac. We find that by increasing the feedback between ECM and Rho, persistent cells become oscillatory. Further increasing the feedback parameter increases the frequency of the oscillations. By extending our model to static 2D domains, we found a new phenotype within the oscillatory regime: spiraling of Rac around the cell edge, that may turn around to spiral in the opposite direction. By simulating cell edge deformations, we also find intermediate phenotypes: cells that oscillate with irregular frequencies, and persistent cells that occasionally attempt but fail to re-polarize, and remain persistent. We also find cells that create and retract protrusions  at random new locations (the random phenotype). Such behaviours cannot be detected in the 1D (or even static 2D) simulations.  Finally, we observed that spontaneous pattern formation in motile cells occurred in parameter regimes where static simulation predicted uniform solutions.

We found that changes in  biophysical properties such as cell stiffness can affect the internal patterning dynamics. So, the same signaling network and rates, but slightly different biophysical parameters result in very different overall cell behaviour. This suggests that cell biophysics is an important factor in cell migratory behaviour that should be more closely examined in the context of intracellular signaling. See \cite{Zmurchok2020b} for an example of this type.

We developed our Models II and III based on integrin bond dynamics modeled by Novikova and Storm \cite{Novikova2013}. That is, adhesions are depicted as clusters of catch-slip bonds that stabilize with intermediate force, and break beyond some force threshold. Many other models for catch-bond dynamics exist, with slightly different assumptions and greater complexity \cite{mackay2020}. The advantage of the version we used is in its simplicity and the fact that it was fit to experimental data for force-lifetime curves of $\alpha 5-\beta 1$ integrin bonds \cite{Novikova2013}. This type of integrin is expressed by many cells and binds to fibronectin. 
Future data for force-lifetime curves of other cell types, other ECMs, annd integrin types could be used to refine parameters and equations. Other interesting extensions include modeling adhesions as mixtures of different integrin bonds (e.g. combination of slip and catch-slip bonds \cite{Novikova2021}). Such model extensions could help to further explain the response of intracellular signaling to structural and spatial variations in the ECM and integrin expression of cells \cite{Danen2002,danen2005}.

%\LEK{\textbf{Relevance to real cells}}

Real cells exhibit a wide range of dynamic behaviour. Examples include  actin waves, lamellipodial ruffles, random filopodial protrusions, circular cell motility, oscillations in situ, and many more. The underlying mechanisms are often unclear, and the distribution of signaling proteins in such dynamics is only rarely characterized. Here, we have shown that various exotic dynamics are inherent in even simple subsets of signaling networks. Such internal dynamics can affect a cell's ability to migrate as well as its invasive potential. %Those melanoma cells with persistent phenotype are seen in our simulations to migrate over long distances.

Our results also contribute to further research on ECM-cell signaling. We have expanded on the work of \cite{JSpark2017} with more highly resolved models, finding overall agreement. We are able to actually simulate the class of random polarity and motility that previous models did not resolve. Finally, the specific details of the model could be altered to generalized to other sources of feedback or to other hypotheses about the influence of ECM signaling.

%\EGR{ \textbf{Limitations}}

Our models have several limitations that result from simplifications or assumptions we made. First, we have adopted the hypothesis from \cite{JSpark2017} that the Rac-Rho module accounts for major aspects of melanoma cell migratory phenotype control. This patently ignores thousands of components that provide additional feedback, input, or fine-tuning. We also directly linked Rac and Rho to cell edge dynamics, ignoring the delays in recruitment and assembly of actin, as well as the activation of myosin motors. We have also ignored the inhomogeneities of the cell and its thickness in the third dimension. All these are common simplifications. 

Another simplification is our expression for the traction forces of the cell. We employed a simple phenomenological relation between traction force and levels of Rac/Rho in the cell. More detailed force models, as in \cite{Rens2019,Rens2020} could be used to increase accuracy and relate other biophysical parameters (stiffness, membrane tension) to the forces that are applied on the integrins.

Results in the paper provide examples of possible outcomes. The kymographs for moving cells are for one realization only, and tend to vary somewhat from one simulation to another. Both the initial conditions and also the stochastic nature of the CPM affect the given outcome. As noted, there are regions of parameter space in which multiple behaviours coexist, and while we display examples of this type, we made no attempt to fully characterize these.

%\LEK{\textbf{Future prospects}}
With the basic behaviour mapped out into regimes in parameter space, our next investigation (ongoing) is to understand implications to real cells and compare to new experiments. The full model can now be used to study the effect of local variations in the ECM (or post density arrays) that lead to directed cell motility. The ability of cells to migrate towards topographical cues (topotaxis) as well as the effect of cell and substrate stiffness provide a rich new set of phenomena to explore computationally. 
Cell migration also affects the deposition and degradation of ECM \cite{Park2020}, providing yet another set of phenomena for investigation in the future.

\section{Acknowledgements}

While at UBC, EGR was funded by a Natural Sciences and Engineering Research Council of Canada (NSERC) Discovery Grant 41870 to LEK. We thank Andreas Buttensch\"{o}n for his FastVector library and general help with optimizing the CPM code written by EGR. We are grateful to Lutz Brusch, J\"{o}rn Starru{\ss} and Andreas Deutsch for the open software Morpheus, and to Bard Ermentrout for XPPauto. We are grateful to Zachary Pellegrin for his development of a CPM reaction-diffusion solver. We thank Andre Levchenko and JinSeok Park for early discussions about this project. We thank Andreas Buttensch\"{o}n for his FastVector library and general help with building the CPM code.

\section{Appendix}

\subsection{Full Model Equations}
For convenience, we gather full model equations in this section.

%\edits{[ADD THE FULL SET OF PARAMETERS FOR EACH MODEL SO THEY CAN BE REPRODUCED.]}

\subsubsection{The well-mixed Rac-Rho system} 

\begin{subequations}
\label{eq:WellMixedRacRho2}
\begin{eqnarray}
\frac{d R}{dt} &=& \frac{b_R }{1+\rho_a^3}(R_T-R) - \delta R,\\
\frac{d \rho}{dt} &=& \frac{b_\rho }{1+R_a^3}(\rho_T-\rho) - \delta \rho .
\end{eqnarray}
\end{subequations}

\subsubsection{The well-mixed Rac-Rho-ECM Model I}

\begin{subequations}
\label{eq:LEK_ModelI}
\begin{align}
\frac{dR}{dt}&=\frac{b_R}{(1+\rho^3)}(R_T-R)-\delta R, %+D\Delta R
\\
\frac{d\rho}{dt}&=\left(k_E+ \gamma_E \frac{E^m}{E_0^m+E^m}\right)\left( \frac{1}{1+R^3}\right) (\rho_T-\rho)-\delta \rho, 
%+D\Delta \rho 
\\
% \frac{dR_I}{dt}&=-\frac{b_R}{(1+\rho^3)}R_I+\delta R +D_I\Delta R_I\\   
%\frac{d\rho_I}{dt}&= -\gamma_E f_{E} \frac{\rho_I}{1+R^3}+\rho +D_I\Delta \rho_I\\
\frac{d E}{dt}&= \epsilon \left[\left(k_R+\gamma_R \frac{R^n}{R_0^n+R^n}\right)-E\left(k_\rho+\gamma_\rho  \frac{\rho^n}{\rho_0^n+\rho^n}\right)\right].
%\mbox{where    } & f_{E}= \frac{E^m}{(E_0^m+E^m)},\quad f_{R}=\frac{R^n}{(R_0^n+R^n)} , \quad f_{\rho}= \frac{\rho^n}{(\rho_0^n+\rho^n)}
\end{align} 
\end{subequations}

\subsubsection{Spatially distributed Rac-Rho-ECM Model I}

In each of the spatial model variants, we can no longer eliminate the total GTPase, since now conservation leads to an integral constraint,
\[
R_T = \int_\Omega (R + R_I) d \Omega = \mbox{constant},
\]
and similarly for Rho (where $\Omega$ is the domain representing the cell). Hence we keep the full PDEs for active and inactive Rac, and active and inactive Rho.

\begin{subequations}
\label{eq:LEK_PDEs1D}
\begin{align}
\frac{dR}{dt}&=\frac{b_R}{(1+\rho^m)}R_I-\delta R +D\Delta R,
\quad  \frac{dR_I}{dt}=-\frac{b_R}{(1+\rho^m)}R_I+\delta R +D_I\Delta R_I,
\\
\frac{d\rho}{dt}&=  \frac{b_\rho(E)}{1+R^m}\rho_I-\delta \rho \, +D\Delta \rho, \quad 
\frac{d\rho_I}{dt}= - \frac{b_\rho(E)}{1+R^m}\rho_I+\delta \rho \, \, +D_I\Delta \rho_I,\\
\frac{dE}{dt}&= \epsilon(a(R)-d(\rho) E),
\end{align}
where the functions appearing above are
\begin{align}
b_\rho(E)&= 
k_E+ \gamma_E \frac{E^m}{(E_0^m+E^m)},\\
a(R)&=K+\gamma_R \frac{R^n}{(R_0^n+R^n)},\\
d(\rho)&= k_P+ \gamma_\rho  \frac{\rho^n}{(\rho_0^n+\rho^n)}.
\end{align}
\end{subequations}
We assume that $D_R=D_\rho=D$, since GTPases have similar sizes. The active form diffuses much slower than the inactive form, so $D<<D_I$.

\subsubsection{Spatially distributed Models II and III}
We use the same Rac -Rho equations as in \eqref{eq:LEK_PDEs1D}a,b, d,  with a new PDE for the ECM variable:

\begin{equation}\label{eq:ECM_ModelII_IIIApp}
\frac{dE}{dt}=\epsilon \left(K(E_t-E) -  d(F,E) E \right).
\end{equation}
where the force on an integrin bond is
\begin{equation}\label{eq:forceonbond}
  F(\rho,R)=\mbox{max}\left(0,\frac{\beta_\rho \rho}{(1+\rho)}- \frac{\beta_R R}{(1+R)} \right).
\end{equation}

For the slip-bond (Model II) we use
\begin{equation}\label{eq:db_slipApp}
d(F,E)= k_0\exp \left(\frac{F}{p(E+E_s)}\right).
\end{equation}
whereas for the slip-catch bond (Model III) we use
\beq
\label{eq:dbCatchSplipApp}
d(F,E) = k_0\exp \left(\frac{F}{p(E+E_s)}\right)+k_{0c}\exp \left(-\frac{F}{p(E+E_s)}\right).
\eeq

Altogether, the model structure for the active forms in Models II and III is
\begin{subequations}
\label{eq:LEK_PDEs1DMod2&3}
\begin{align}
\frac{dR}{dt}&=\frac{b_R}{(1+\rho^m)}R_I-\delta R +D\Delta R,
%\quad  \frac{dR_I}{dt}=-\frac{b_R}{(1+\rho^m)}R_I+\delta R +D_I\Delta R_I,
\\
\frac{d\rho}{dt}&= \left(k_E+ \gamma_E \frac{E^m}{(E_0^m+E^m)}\right) \cdot \left(\frac{1}{1+R^m}\right) \rho_I-\delta \rho +D\Delta \rho, 
%\quad \frac{d\rho_I}{dt}= -b_\rho(E) \frac{\rho_I}{1+R^m}+\rho +D_I\Delta \rho_I,
\\
\frac{dE}{dt}&=\epsilon \left(K(E_t-E) -  d(F(R,\rho),E) E \right),
\end{align}
\end{subequations}
together with two additional PDEs for the inactive GTPases, $R_I, \rho_I$, as before in \eqref{eq:LEK_PDEs1D}.

\subsection{Parameter values}

We employed nondimensionalized parameters (see Table \ref{tab:pars}). We can define a spatial and temporal scale, by comparing with melanoma cells in \cite{park2016,JSpark2017}. These cells have a diameter of roughly 30 $\mu$m \cite{JSpark2017} and our CPM cells have a diameter of 45 pixels. So 1 CPM pixel can be assigned a width of 2/3 $\mu$m. Given that one CPM pixel is 0.075 in non-dimensional units, our spatial scale is then given by $L_0 \approx 9 \mu$m. We estimate the speed of persistent melanoma cells in \cite{park2016} as 
0.1 $\mu$m/min, similar to  melanoma migration speeds reported elsewhere, of 5-40 $\mu$m/hr \cite{Li2017,Howard2013}.
The simulated persistent cell in Fig.~\ref{fig:threephenotypes-v1}A has a speed of roughly 0.02 $\mu$m/MCS. During one MCS, we perform 1 timestep of the reaction-diffusion equations.  So, one MCS is associated with $t_0$=12 s.  These temporal and spatial scale result in a diffusion rate of $\frac{L_0^2}{t_0} \approx 6.5\mu$m$^2$/s, for the inactive form of the GTPases. This gives realistic values for our parameters, and they can be adapted to link with different types of cells and conditions.

\subsection{Kymographs in 2D}
In both static and deforming 2D, we produce kymographs by tracking the Rac level on the cell edge. We identified the cell edge using the edge-detection function bwboundaries in Matlab. To align the Rac front to the left of the kymograph, we first identify the location of highest Rac relative to the cell centre (See methods in \cite{JSpark2017}) and rotate the image by that angle.

\subsection{Implementation details, Morpheus}

\subsubsection{Morpheus software and files}

\begin{table}[]
    \centering
    \begin{tabular}{|l|l|l|}\hline
 Results  & Figure  &Morpheus xml file \\
 \hline
 Model I in 1D& Fig.~\ref{fig:ModelI_K-gammeE-bR}&
 Model1RacRhoECMPDEsIn1D \\
 Model II in 1D& Fig.~\ref{fig:ModelII_Et-gammeE-bR}& Model2SlipBondRacRhoECMPDEsIn1D \\
 Model III in 1D&Fig.~\ref{fig:Model3:Et-gammeE-bR-catch}& Model3CatchSlipBondRacRhoECMPDEsIn1D\\
 One cell, WM Model I&Fig.~\ref{fig:OneLamelCellSize}&  OneLamelShape\\ 
2Lam  Model I  &Fig.~\ref{fig:TwoLamelMorpheusModelResults} &Melanoma2LamelCPMRac-Rho-ECMWellMixed \\
 \hline
    \end{tabular}
    \caption{A list of the Morpheus files used to prepare a number of figures. WM= well-mixed, 2Lam = two -lamellipod.}
    \label{tab:MorpheusFilesforfigs}
\end{table}

Morpheus is an open-source software package for multiscale simulations of single and interacting cells \cite{Starruss2014}. Morpheus solves differential equations, and simulates the cellular-Potts model to depict cells. Morpheus produces small .xml files that allow users to exactly reproduce simulation results.

\subsubsection{Cell shapes with well-mixed internal Rac-Rho-ECM}

For the cell shape simulation with well-mixed internal variables, we linked the internal chemistry to the target area $A_0$ in the Hamiltonian,
\begin{equation}
H = \lambda_A (A-A_0)^2 + \dots
\end{equation}

\textbf{One cell compartment:} For the single-cell compartment shown in Figure~\ref{fig:OneLamelCellSize}, we used the well-mixed Rac-Rho-ECM Model I and assumed that 
 \[
 A_0=400 + 1000\cdot R-100 \cdot \rho.
 \]

\begin{figure}
    \centering
    \includegraphics[scale=0.5]{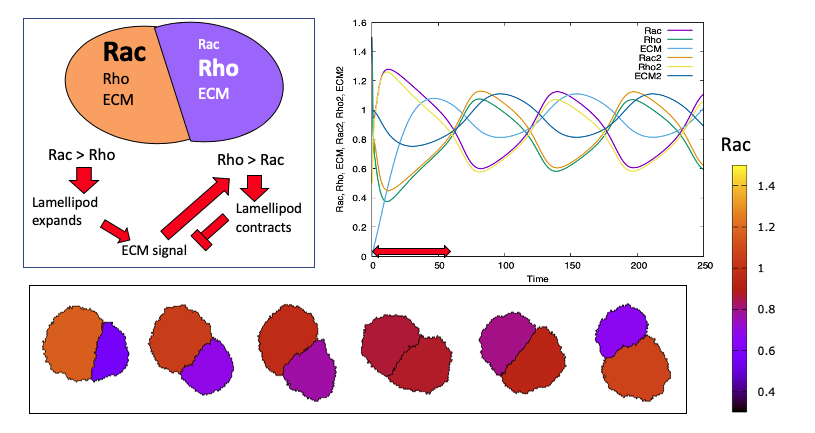}
    \caption{The two-lamellipod model. Top left: a schematic diagram indicating how the various interactions are modeled based on Eqs.~\eqref{eq:LEK_ModelI} of Model I. %Eqs.~\eqref{eq:WellMixedRacRho}-\eqref{eqn:ECM_01}. 
    Top right: A plot of the variables over time in each of the two lamellipods, created with Morpheus file Melanoma2LamelCPMRac-Rho-ECMWellMixed.xml. The red arrow spans the time-frame displayed across the bottom row. Bottom: Morpheus simulation showing the ``shape'' of a cell consisting of the two compartments. The color scheme gives the Rac level in each compartment. The areas of the ``front'' and ``back'' lamellipods evolve as the levels of GTPases oscillates. First Rac is high on the left, and then it shifts to being high on the right. Produced with Morpheus file Melanoma2LamelCPMRac-Rho-ECMWellMixed.xml}
    \label{fig:TwoLamelMorpheusModelResults}
\end{figure}

%\subsubsection{Polarity flips in a Two-lamellipod well-mixed Model I CPM cell}
\textbf{Two cell compartments:} We also ran Morpheus simulations in which a single cell was depicted by by a pair of geometric objects as in \cite{Schnyder2017}. This is directly motivated by \cite{Holmes2017} where two ``lamellipods'' compete for ``frontnes''. As example of this type is shown in Figure~\ref{fig:TwoLamelMorpheusModelResults}. 
The levels of active Rac and Rho, $R_1, R_2, \rho_1, \rho_2$ in each lamellipod, and the ECM contact area of each, $E_1, E_2$ are given by the system
\eqref{eq:LEK_ModelI}
but the compartments are coupled by  assuming that,
\begin{equation}
\label{eq:TwoLamelCons}
 R_T= R_I+R_1+R_2, \quad \rho_T= \rho_I+\rho_1+\rho_2.   
\end{equation}
Since inactive GTPases $R_I, \rho_I$ diffuse rapidly, it is reasonable to assume that their levels are the same in both compartments. Hence, we can eliminate $R_I, \rho_I$ from the well-mixed Model I ODE system.
We assumed that the target area of each compartment was given by
\[
A_0=400 + 1500R-300\rho.
\]
The adhesive energy between the two compartments is kept purposely low to ensure that they adhere to form a single ``cell''. 
As shown in Figure~\ref{fig:TwoLamelMorpheusModelResults}, the two lamellipods oscillate in antiphase manner, with high Rac in one corresponding to low Rac in the other.

\subsection{Implementation details, custom-built CPM simulations}

The CPM is an energy-based cell shape computation, where the energy is represented by a Hamiltonian \cite{Graner1992,Magno2015}.
In a single-cell model, the Hamiltonian has typical terms that include the following:

\begin{equation}
H = \lambda_A (A-A_0)^2 + J P + \lambda_P (P-P_0)^2 \label{Eq:H}.
\end{equation}
These terms include energetic penalties for cell area that deviates from a ``rest area''  $A_0$, for perimeter deviation from the value $P_0$. There is also an energy associated with cell-medium contact, depicted by the term $JP$. 

A cell is represented by some set of contiguous pixels (in 2D for our paper).
During each Monte-Carlo step (MCS), the cell attempts a number of protrusions/retractions at pixels along its boundary, and accepts all attempts that decrease the Hamiltonian.  Otherwise, the probability of accepting a change that increases $H$ is given by a Boltzman probability distribution that depends on the ratio of energy gain to some energy fluctuation (``thermal energy'' term). More information about the CPM formalism and a comparison with other methods of simulating cells is provided in \cite{Magno2015,Rens2019}.

\subsubsection{Imposed cell elongation}

We added cell elongation to avoid circular motion and to keep cell trajectory straight with a persistent front. (See \cite{camley2017} for circular motion in a simple GTPase cell-shape model.)
To impose cell elongation we add an additional term to the Hamiltonian
\begin{equation}
    H_e = \lambda_e \cdot (100 \cdot \xi - \xi_0)^2,
\end{equation}
where $\xi$ is the eccentricity of the cell. To approximate the eccentricity \cite{Zajac2003}, we used the inertia tensors $I(S)$ of the cells, defined as 
\begin{equation}
I=\left(
\begin{matrix}
\sum(y-c_y)^2 & -\sum(x-c_x)(y-c_y) \\
-\sum (x-c_x)(y-c_y) & \sum (x-c_x)^2
\end{matrix}\right),
\end{equation}
where the sum goes over all the coordinates within the cell, and $(c_x,c_y)$ is the center of mass of the cell.
The two eigenvalues $a,b$, with $a>b$, of this matrix approximate the relative sizes of the long and short axes of the cell. The eccentricity of a cell is defined as $\xi = \sqrt{1-\left({\frac{b}{a}}\right)^2}$. A circular cell has $\xi \approx 0$, while an elongated cell has $\xi \approx 1$. This elongation constraint deviates from the typical cell length constraint (see e.g. \cite{Zajac2003}), which only affects the long and not the short axis. Using the old constraint, a cell tended to round up.
%easily became round as the short axis would become almost as big as the long axis. 
Our new elongation constraint prevented this and assured that the cell long and short axes were significantly different.
%imposes the shorter axis to be considerately shorter than the long axis.

\subsubsection{Forces due to Rac-Rho activities}
Technically, we implement the Rac-Rho dependent edge forces through the CPM Hamiltonian, as done, e.g., in \cite{Maree-06,Rens2019}. That is, we modify the local change in the Hamiltonian (equivalent to a local force) as follows 

\begin{eqnarray}
\Delta H = \beta_G \cdot \begin{cases}
\beta_R \frac{R_a}{1+R_a} - \beta_P \frac{\rho_a}{1+\rho_a}, & \text{if retraction}, \\
-\beta_R \frac{R_a}{1+R_a} + \beta_P \frac{\rho_a}{1+\rho_a}, & \text{if extension}. 
\end{cases}
\end{eqnarray}
Here, $\beta_G$ sets the relative change in energy ($\approx$ force) contribution of the GTPases compared to terms in the general Hamiltonian.

\subsubsection{Forces needed to overcome adhesion bonds:}

To make the cells less motile when their ECM attachment is higher, we implement an additional ECM-dependent yield energy, $H_0$, (a kind of dissipative energy) that the cell needs to overcome to make a movement. To do so, we used the probability function

\begin{equation}
P(\Delta H) = \begin{cases}
1 & \textrm{if}\;\Delta H + H_0 <0 \\
e^{-(\Delta H+ H_0)/T} & \textrm{if}\;\Delta H + H_0\geq 0,
\end{cases}
\end{equation}
where $H_0$ depends on the cell-ECM adhesivity as follows:
\begin{equation}
    H_0 = \begin{cases}
    0 & \text{if extension} \\
    \lambda_N E & \text{if retraction}.
    \end{cases}
\end{equation}

\subsubsection{Initial conditions}

%In the 1D simulations, we set \edits{FILL IN WHAT'S MISSING.}

For the 2D simulations, we use a circular domain of diameter $L$=3. We imposed uniform noise in active Rac throughout the cell with a %We set the 
maximal level of 
%uniform noise 
0.8.
%, corresponding to a similar total Rac if we would have activated the left 10\% of a 2D region of size 9. 

In 1D and static 2D simulations, we used grid size $dx=0.1$ to solve the PDEs, but for the deforming cells, we increased the resolution by setting $dx=0.075$.

%\edits{MOVED HERE FROM CPM SECTION}
We initiate the simulation with a cell with uniform adhesion: $N(\vec{x})=N_u$, where $N_u$ is some small value, chosen here as 50. Whenever the cell protrudes, we assume it binds to the ECM and set $N(\vec{x})=N_0=50$. After a cell retracts, the ECM attachment is set to zero.

\subsubsection{Redistribution of fields after cell moves}
After a protrusion, the newly created cell site obtains the GTPase levels of the pixel that was ``copied'' to the target site. After a retraction, the GTPase level formerly at a cell site is set to zero. After each MCS, we normalize the GTPase levels (over the whole cell) to ensure conservation of the total GTPase of each type. %After a retraction, the adhesion density $E$ is set to zero and after a protrusion it is set to $E_n=50$.

\newpage

\begin{table}[]
    \centering
    \begin{tabular}{|l|l|l|}
    \hline
            parameter & description & value   \\ \hline \hline
    \textbf{space/time}  & &   \\
    $dx$ & lattice site (pixel) width & 0.075    \\  
    $dt$ & PDE integration time-step & 0.00125   \\  
    $n_i$ & number of PDE integration time steps & 800   \\
    $N$ & number of Monte Carlo Steps in CPM & 1000   \\
    \hline \hline
    \textbf{CPM}  & &   \\
        $A_0$ & target area & 1600   \\
        $\lambda_A$ & strength of area constraint & 0.5  \\
        $P_0$ & target perimeter & 200   \\
        $\lambda_P$ & strength of perimeter constraint & 12   \\
        $e_0$ & target eccentricity & 90   \\
        $\lambda_e$ & strength of eccentricity constraint & 20  \\
        $J$ & adhesive energy & 40000   \\
        $n_r$ & neighbourhood radius for $J \cdot P$ & 3   \\
        $\xi(r)$ & scaling factor for $J \cdot P$ & 18   \\
        $\beta_G$ & GTPase force strength & 0.5    \\
        $\lambda_N$ & adhesion strength & 0.3   \\
        $T$ & cellular temperature & 200   \\ 
        \hline \hline
        \textbf{Rac/Rho} & &   \\
        %$D_{\rho_i}$ 
        $D_I$
        & diffusion of inactive Rac and Rho & 1   \\
        %$D_{\rho_a}$
        $D$
        & diffusion of active Rac and Rho & 0.1    \\
    %    $\delta_P$ & Rho inactivation rate &  1 &  \\
     %   $D_{R_i}$ & diffusion of inactive Rac & 1 &  \\
      %  $D_{R_a}$ & diffusion of active Rac & 0.1 &  \\
    %    $\delta_R$ 
    $\delta$
        & Rac, Rho inactivation rate & 1   \\
        $b_R$ & Rac activation rate & 2   \\
         \hline \hline
        \textbf{ECM-Rho coupling} & &   \\
        $k_E$ & basal Rho activation rate & 2   \\
        $\beta_R$ & protrusive force of Rac & 1200 or 1000    \\
        $\beta_P$ & contractile force of Rho & 1600   \\
        $E_0$ & half-max adhesion for Rho activation & 300   \\
        $\gamma_E$ & adhesion dependent Rho activation rate & 4   \\
         \hline \hline
        \textbf{Adhesion model} & &   \\
        $r$ & integrin binding rate & 10   \\
        $E_T$ & maximum adhesion size & 1000   \\
        $k_0$ & integrin unbinding rate due to slip & 5 or 0.0004    \\
        $k_{0c}$ & integrin unbinding rate due to catch & 0 or 55   \\
        $\epsilon$ & time-scale  & 0.001   \\
        $p_s$ & scaling factor force & 0.35 or 0.08   \\
        $E_s$ & small adhesion for scaling & 100   \\
        $E_n$ & nascent adhesion at new protrusion & 50   \\
    \hline
    \end{tabular}
    \caption{Dimensionless parameters for the CPM simulations. The value options indicated by "value1 or value2" signify the different values chosen for Model II (slip) or Model III (catch-slip).}
    \label{tab:pars}
\end{table}

%%%%%%%%%%%%%%%%%%%

\clearpage

\bibliographystyle{vancouver}
\bibliography{GTPasebib}

\clearpage

\end{document}